\documentclass[prb,reprint,amsmath,amssymb,showpacs,superscriptaddress,floatfix,longbibliography]{revtex4-1}

\usepackage{epsfig,mathrsfs,color,latexsym,subfigure,marginnote,gensymb,}
\usepackage{graphicx}
\usepackage{makecell}
\renewcommand{\BibitemShut}[1]{}

\begin{document}
\title{Role of disconnections in mobility of the austenite-ferrite inter-phase boundary in Fe}
\author{Pawan Kumar Tripathi}
\affiliation{Dept. of Materials Science and Engineering, Indian Institute of Technology Kanpur, Kanpur 208016, India}
\author{Sumit Kumar Maurya}
\affiliation{Dept. of Materials Science and Engineering, Indian Institute of Technology Kanpur, Kanpur 208016, India}
\author{Somnath Bhowmick}
\email{bsomnath@iitk.ac.in}
\affiliation{Dept. of Materials Science and Engineering, Indian Institute of Technology Kanpur, Kanpur 208016, India}
\date{\today}

\begin{abstract}
Austenite ($\gamma$-Fe, face centered cubic (FCC)) to ferrite ($\alpha$-Fe, body centered cubic (BCC)) phase transformation in steel is of great significance from the point of view of industrial applications. In this work, using classical molecular dynamics simulations, we study the atomistic mechanisms involved during the growth of the ferrite phase embedded in an austenite phase. We find that the disconnections present at the inter-phase boundary assist in growth of the ferrite phase. Relatively small interface velocities (1.19 - 4.67 m/s) confirm a phase change via massive transformation mechanism. Boundary mobilities obtained in a temperature range of 1000 to 1400 K show an Arrhenius behavior, with activation energies ranging from 30 - 40 kJ/mol.
\end{abstract}

%
%
\maketitle

\section{Introduction}
\label{intro}
During the process of iron and steel making, as molten Fe is cooled, first it solidifies to $\delta$ (BCC) allotrope of iron at a temperature of 1811 K. This is followed by solid-solid phase transformations, initially from $\delta$-Fe to $\gamma$-Fe (FCC) at 1667 K and finally from $\gamma$-Fe to $\alpha$-Fe (BCC) at 1185 K. The latter is very important, because the microstructure and mechanical properties of Fe-alloys are governed by the amount of austenite ($\gamma$-Fe) and ferrite ($\alpha$-Fe) present after the transition. Being a very complex process, governed by several extrinsic (composition, rate of cooling etc.) and intrinsic (nucleation, inter-phase and grain boundary mobility, relative orientation of the two phases etc.) factors, the atomistic mechanisms involved during the phase transition are not clearly understood yet. Based on several experimental studies, it has been established that the nature of the transition is either martensitic or massive.\cite{roitburd, hillert}  The former is a diffusion-less transformation, which takes place via a coordinated movement of atoms by a distance less than the inter-atomic spacing. On the other hand, massive transformation occurs via nucleation and growth of the ferrite phase at the expense of the austenite phase, driven by Gibbs free energy change. 
 
In order to describe the kinetics of the $\gamma$-$\alpha$ transformation, mainly two types of models have been proposed in the literature; diffusion controlled growth model\cite{zener} and interface controlled growth model.\cite{christian} In reality, transformations are mixed in nature, starting as an interface controlled process and following the initial stages of nucleation and growth, a relatively slow diffusion controlled process takes over.\cite{sietsma_z,krielaart} The interface controlled phase transformation is characterized in terms of intrinsic mobility of the inter-phase boundary and values ranging from $10^{-6}$ to $10^{-9}$ m-mol/(J-s) have so far been reported in the literature.\cite{hillert,gamsjager,krielaart}  Boundary mobilities are also known to show an Arrhenius behavior, with activation energy reported to be $\approx 140$ kJ/mol.\cite{hillert,gamsjager,krielaart}

Since the nucleation and growth of the ferrite phase starts at the $\gamma$-$\alpha$ inter-phase boundary, orientations of the two phases at the interface play a crucial role in transformation. Several orientation relationships (OR) between the FCC and BCC phase have so far been proposed in the literature. This includes Bain,\cite{bain} Nishiyama-Wasserman (NW),\cite{nw1934} Kurdjumov-Sachs (KS),\cite{Kurdjumow1930} Greninger-Troiano (GT)\cite{gt} and Pitsch.\cite{pitsch} Other than the Bain and Pitsch, interface is formed between the two closest packed planes of the two phases, i.e., (111)$_{\rm{FCC}}$ $\parallel$ (110)$_{\rm{BCC}}$. Pitsch OR is exactly opposite to this, with (111)$_{\rm{BCC}}$ $\parallel$ (110)$_{\rm{FCC}}$. In case of Bain OR, interface is formed between the (001) plane of both the phases. Among all the ORs, NW and KS are more often reported in case of iron and steel.\cite{fukino}

Because of its length and time scale, interface controlled $\gamma$-$\alpha$ phase transformation can be investigated by atomistic calculations\cite{dudarev,katanin2016,leonov2011,leonov2012} and several studies related to massive and martensitic transformations based on classical molecular dynamics simulations have been reported so far.\cite{ou,ou_w,bos,song1,song2,tateyama,urbassek,sietsma_acta18} During martensitic transformations, interface velocities are found to be very high, ranging between 200-700 m/s in case of Bain and KS ORs at different temperatures.\cite{bos} On the other hand, much smaller interface velocities (0.7-3.4 m/s) are obtained in case of massive transformation, as reported for a $\gamma-\alpha$ interface of NW type.\cite{song1,song2} A comparison between the NW and KS ORs reveals planar and needle like growth of the ferrite phase taking place at the respective interfaces, the former being ten times slower than the latter.\cite{tateyama} Bi-directional transformations are also reported in case of NW orientation, with significant difference of interface velocity between the $\gamma-\alpha$ (24 m/s) phase change and vice versa (240 m/s).\cite{urbassek}  

Interestingly, in many of the computational studies mentioned above, some kind of defect (like a free surface, stacking faults, twin boundaries, steps present at the $\gamma-\alpha$ interface etc.) is present in the initial structure, which assists the phase transformation. Motivated by this, we focus on a particular type of defect, known as disconnections. This a type of interfacial defect having both dislocation and step-like character.\cite{hirth1994, hirth1996acta,hirth_review} Disconnections are reported to be observed at the inter-phase boundaries of several ferrous and non-ferrous materials.\cite{zhang, maresca_acta18, Ti_ref2, howe, hall} They are also reported to play important role during the phase transformation.\cite{hirth1996acta,hirth_review,aaronson,shiflet,moritani} In this paper we investigate role of disconnections during the austenite to ferrite transformation in pure-Fe, using classical molecular dynamics simulations. Our study clearly shows that the  disconnections located at the austenite-ferrite interface facilitate the growth of the $\alpha$-Fe phase. We also calculate the velocity and mobility of the $\alpha-\gamma$ interface and the values suggest a massive transformation from $\gamma$-Fe to $\alpha$-Fe.
   
The paper is organized as follows: in Sec.~\ref{sd} we discuss the simulation details, which include A) a discussion on interatomic potential, B) calculation of driving force for the phase transformation, C) crystallographic description of the simulation box and D) calculation of interface velocity and mobility. This is followed by a detailed discussion of the main results obtained in this wrok in Sec.~\ref{rd} and the paper is concluded in Sec.~\ref{con}.

\section{Simulation Details}
\label{sd}
\subsection{Interatomic Potential}
\label{IP}
All the calculations are performed using classical molecular dynamics (MD) simulations, as implemented in Large-scale Atomic/Molecular Massively Parallel Simulator (LAMMPS) package.\cite{lammps} Interaction among metal atoms are well approximated by EAM (embedded atom method) potentials. In this work we use the empirical potential developed by Ackland et al.\cite{ackland} Using this potential, calculated values of lattice parameter, cohesive energy, vacancy formation energy and elastic constants are found to be in very good agreement with experimental data, as well as density functional theory (DFT) based predictions [see Table~\ref{T1}]. However, there are two drawbacks of this potential. First, it overestimates the melting point [see Table~\ref{T1}], a fact already reported in the literature.\cite{sunprb2004} Second, the BCC phase remains more stable than FCC up to the melting temperature. In reality, a BCC to FCC phase transition is observed at 1185 K in the experiments, which can not be captured by this empirical potential. Despite these limitations, Ackland potential has been used in numerous MD studies on Fe, including the FCC to BCC phase transition.\cite{song1, song2, ack_ref, ack_ref2}
\begin{table}
\caption{Comparison of physical properties estimated using the empirical potential proposed by Ackland et al.(1997)\cite{ackland} with DFT calculations or experimentally measured values.}
\begin{tabular}{p{4cm}p{2cm}p{2cm}}
\hline
Property & Experiment or DFT &  Ackland et al.(1997) \\
\hline
a (\AA), BCC at T=0 K& $2.855^a$ &2.866 \\
a (\AA), FCC at T=0 K& $3.658 ^a$&3.680\\
T$_m$(K) & $1812^a$ & 2358 \\
E$_{coh}$ (ev/atom)&-4.316$^a$ & -4.316 \\
E$_f^v $(ev/atom) &$1.84^a$ & 1.89\\
C$_{11}$ (GPa) &$242.00^b$&243.39 \\
C$_{12}$ (GPa) & $146.50^b$&145.03\\
C$_{44}$ (GPa) & $112.00^b$& 116.00\\
\hline
\end{tabular}\\
$^a$ Values taken from Mendelev et al.(2003) \cite{mendelev}\\
$^b$ Values taken from Hirth and Lothe (1968) \cite{hl82}
\label{T1}
\end{table} 

\begin{table}
\caption{Driving force for $\gamma$-$\alpha$ transition in iron, calculated using the Ackland potential. }
\begin{tabular}{|c|c|c|c|}
\hline
Temperature(K) & 1000   &  1200  & 1400 \\
\hline
$\Delta G_{\gamma-\alpha}$ (ev/atom) & 0.0198 & 0.0183  & 0.0168 \\
\hline
$\Delta G_{\gamma-\alpha}$ (kJ/mole) & 1.912 & 1.767 & 1.622 \\
\hline
\end{tabular}
\label{T2}
\end{table}

\begin{figure}
\includegraphics[width=1\linewidth]{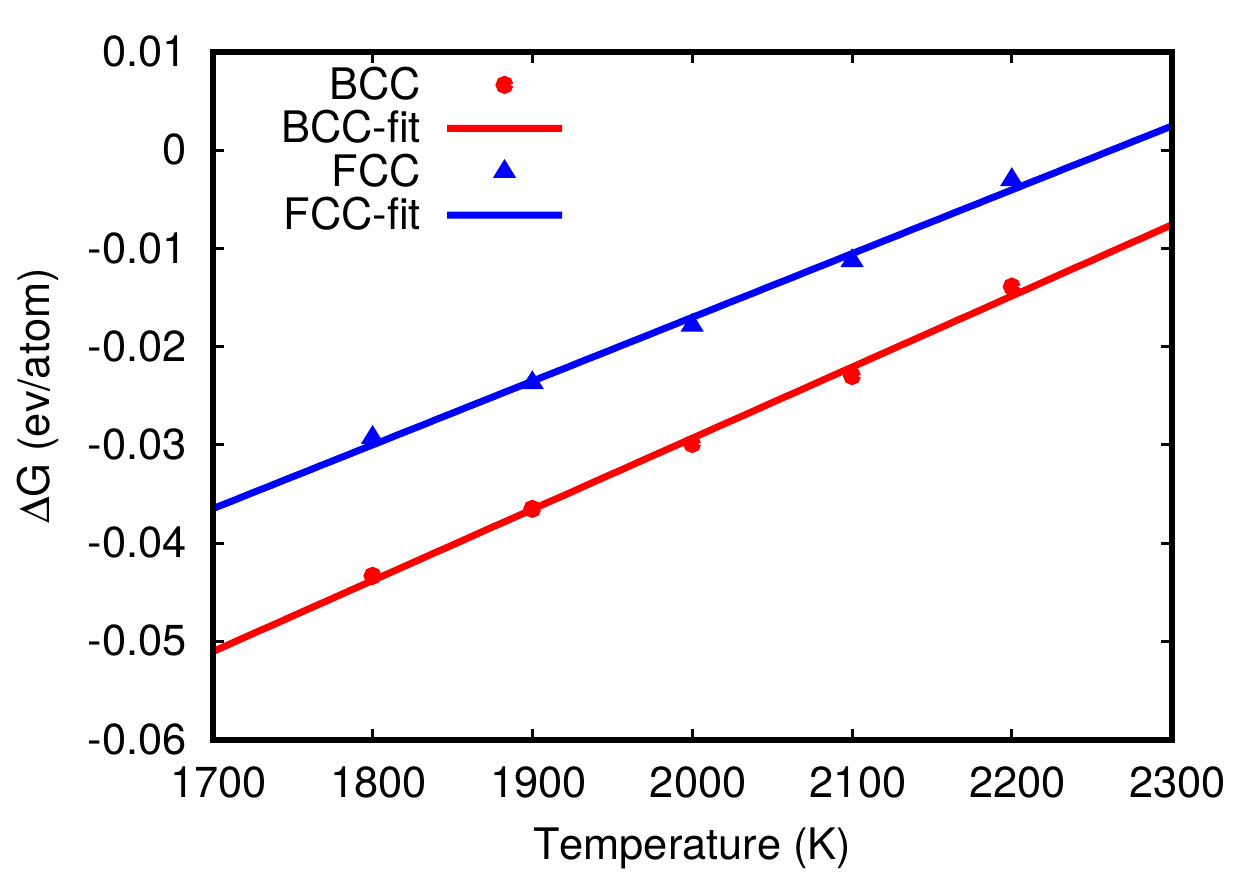}
\caption{Free energy difference between the liquid and solid phase ($\Delta G_{L-S}$) for both $\alpha$-Fe and $\gamma$-Fe. Free energy difference between the solid phases ($\Delta G_{\gamma-\alpha}$) is calculated from the vertical difference between the two lines.}
\label{fig1}
\end{figure}

\subsection{Driving Force for the Phase Transformation}
\label{df}
It is well known that the driving force behind massive transformation is the reduction of the Gibbs free energy ($\Delta G_{\gamma-\alpha}$) as Fe transforms from the austenite to the ferrite phase. In order to calculate $\Delta G_{\gamma-\alpha}$, we first calculate $\Delta G_{L-S}$, the free energy difference between the liquid (L) and solid (S) phase using the Gibbs-Helmholtz equation,
\begin{equation}
\frac{\Delta G_{L-S}}{T}= \int_{T}^{Tm} \frac{H^{S}(T)-H^{L}(T)}{T^{2}} dT,
\label{ghe}
\end{equation}
where $H$ is the enthalpy, which is a function of temperature $T$ and $T_{m}$ is the melting point. The calculation is carried out separately for both $\alpha$ and $\gamma$ solid phases from 1800 to 2200 K at an interval of 100 K [see Fig.~\ref{fig1}]. In case of the solid phase, the simulation box (of size $10\times 10\times 10$) is equilibrated at a given temperature and zero pressure using a NPT ensemble and the enthalpy at that particular temperature ($H^{S}(T)$) is given by the potential energy of the system. On the other hand, in case of the liquid phase (simulation box size $10\times 10\times 10$), first the system is melted at 3000 K and then rapidly cooled to a lower temperature using a NPT ensemble. Rapid cooling ensures that the liquid like structure is maintained even below the melting point and finally the system is subjected to a NVT run (same temperature at which the liquid is cooled) to get the potential energy, equal to the enthalpy of the liquid phase at a given temperature ($H^{L}(T)$). In order to estimate the melting point, we use the coexistence method (developed by Morris and Song \cite{morris}) and $T_m$ for the BCC and FCC phase are found to be 2358 K and 2237 K, respectively. 

Calculated values of $\Delta G_{L-S}$ are illustrated in Fig.~\ref{fig1}, along with a linear fit of $\Delta G_{L-S}$ as a function of temperature, as shown below-
\begin{itemize}
\item FCC: $\Delta G_{L-S} = 6.495\times10^{-5}T - 0.1469$,
\item BCC: $\Delta G_{L-S} = 7.245\times10^{-5}T - 0.1742$.
\end{itemize}
Finally, the free energy difference between the solid phases ($\Delta G_{\gamma-\alpha}$) is obtained from the vertical difference between the two lines at any given temperature. The lines shown in Fig.~\ref{fig1} are extrapolated to get the value of $\Delta G_{\gamma-\alpha}$ at lower temperature, lying in the range of 1000 K to 1400 K. As reported in  Table~\ref{T2}, in the temperature range of 1000 to 1400 K, $\Delta G_{\gamma-\alpha}$ values predicted by EAM potential lie in the range of 1.912 to 1.622 kJ/mole. Although the value dips a bit with increasing temperature, the system is still far from the ferrite to austenite transformation even at 1400 K. This is a well known drawback of the Ackland potential, which prefers ferrite over the austenite phase all the way to the melting temperature. Comparing with the experimental results, it is found that the numerical values of $\Delta G_{\gamma-\alpha}$ obtained from the Ackland potential in this study [see Table~\ref{T2}] are equivalent to the actual free energy difference at a temperature range of roughly 700-750 K.\cite{KAUFMAN1963323} Thus, driving force applied in this study is not completely out of range.

\begin{figure}
\includegraphics[width=0.85\linewidth]{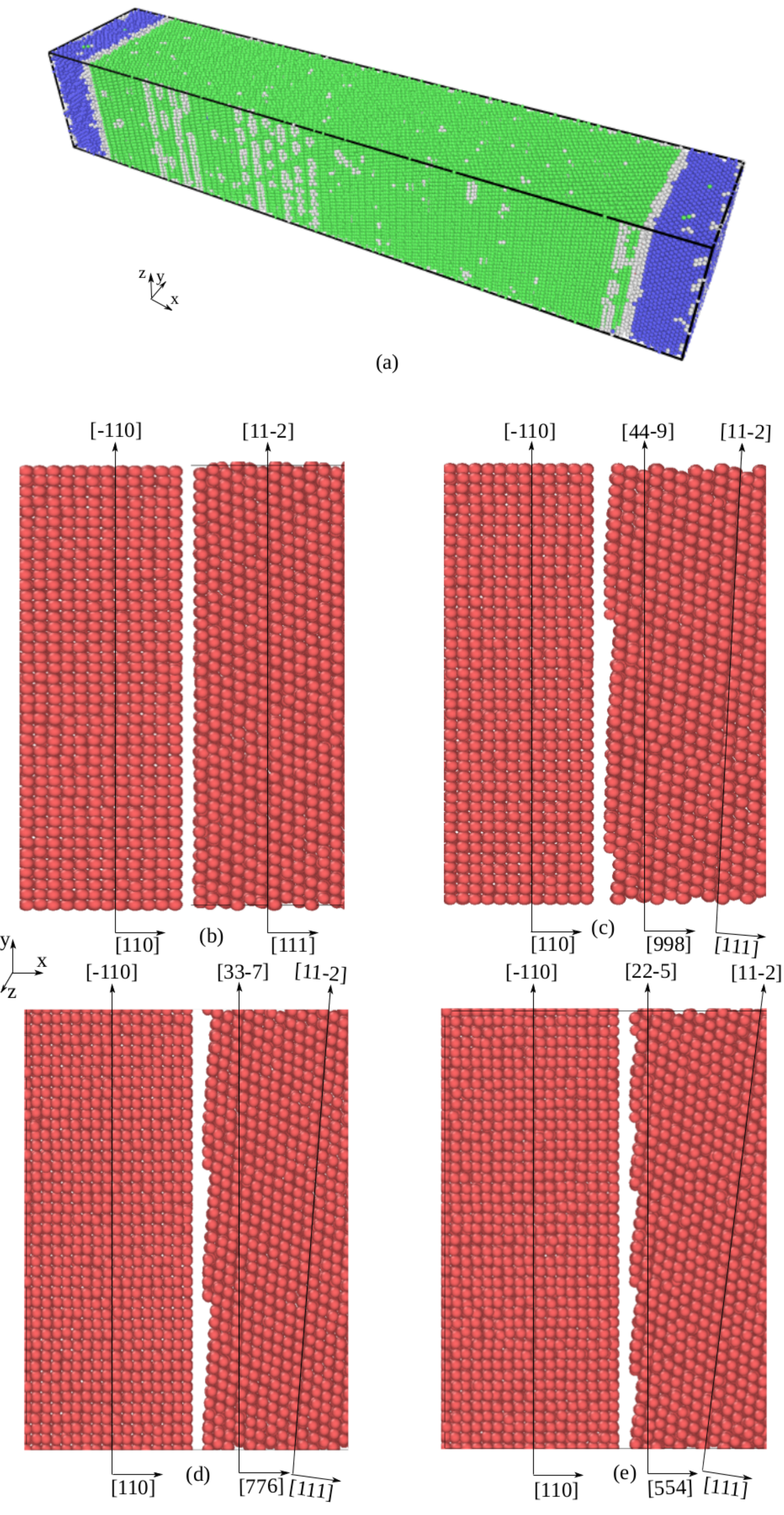}
\caption{(a) Shape of the simulation box after joining the separately equilibrated BCC (blue) and FCC (green) phases, forming a sandwich like BCC-FCC-BCC structure. The figure is prepared using the Ovito software, \cite{ovito} which uses common neighbor analysis to distinguish between the FCC and BCC phase. (b)-(e) Atomic configuration of the BCC-FCC interfaces considered in this work. (b) Atomically flat interface between the (110) plane of BCC and (111) plane of FCC, as per NW OR. Steps or disconnections appear when FCC phase is rotated with respect to the $z$ axis from the ideal NW OR by (c) $3.11^\circ$, (d) $4.04^\circ$ and (e) $5.77^\circ$. Large gaps between the two phases are shown for the sake of visual clarity. In reality, spacing between BCC and FCC region are taken to be the average of BCC (110) and FCC (111) inter-planar distance.}
\label{fig2}
\end{figure}

\begin{figure}
\includegraphics[width=0.75\linewidth]{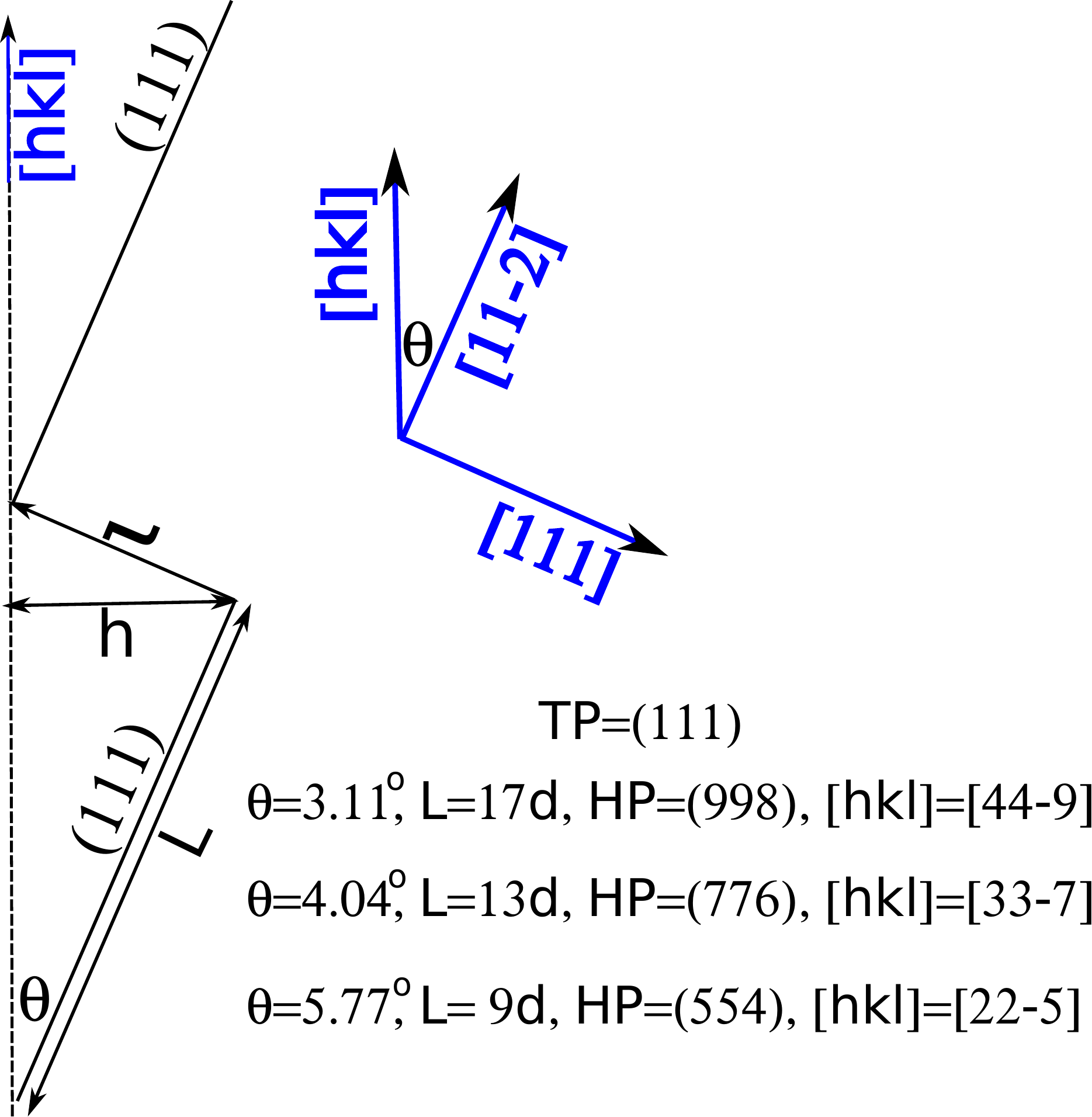}
\caption{A schematic diagram of the disconnections present in the FCC phase, as shown in Fig.~\ref{fig2}(c)-(e). In all the cases, (111) forms the terrace plane (TP), but the habit plane (HP, shown by the dotted line) changes depending on $\theta$, the tilt angle from the ideal NW OR. [hkl] and [11$\bar{2}$] direction lies in the HP and TP, respectively and the angle between the two directions equals $\theta$. $\vec{l}$ is the ledge vector and step height $h$ is measured from the HP. As shown in Fig.~\ref{fig2}, every step is monoatomic. $L$ is the length of the TP, consisting of 17, 13 and 9 atomic rows and $d$ is the diameter of Fe atom.}
\label{fig2a}
\end{figure}

\subsection{Simulation Box Details}
\label{simbox}
Since our goal is to study the FCC to BCC phase transition, we must have both the phases present in the beginning. For this purpose, initially we create a BCC and a FCC box separately. The crystallographic directions parallel to the box edges and sizes of the boxes in terms of number of atomic planes present along a particular direction are reported in Table~\ref{T3}. Note that, the cross-section (yz plane) of both the boxes are chosen such that the area mismatch of the individual interfaces remains less than 0.5\% after we join the two phases at a later stage. A larger mismatch of cross-section between the two phases should be avoided, as it leads to high stresses, which can significantly affect the transformation process. First, both the boxes are equilibrated separately for 2 ns using a NVT ensemble to bring all the atoms in thermal equilibrium. This is followed by a 6 ns run using a NP$_x$T ensemble for the purpose of volume equilibration, without altering the interface area.  

After equilibration, boxes are joined in a sequence of BCC-FCC-BCC, making a sandwich like structure [see Fig.~\ref{fig2}]. After joining the two phases, the simulations box remains fully periodic, having no free surface and there are two BCC-FCC interfaces within the box. Interface is formed parallel to the $yz$ plane and the growth direction (of the BCC phase) is perpendicular to the interface (along the $x$ axis). Since NW is one of the most commonly observed orientation relationship between the BCC and FCC phase, we select this among various possibilities (as mentioned in Sec.~\ref{intro}) to create the austenite-ferrite interface in this work. NW is a semi-coherent interface between the closest packed planes of BCC and FCC phase, described as $(1 1 0)_{\rm{BCC}} \parallel (1 1 1)_{\rm{FCC}}$ and $[0 0 1]_{\rm{BCC}} \parallel [1\bar{1} 0]_{\rm{FCC}}$. In case of ideal NW orientation relationship, an atomically flat interface is created between the BCC (110) and FCC (111) plane [see Fig.~\ref{fig2}(b)]. Keeping the BCC phase fixed, we further tilt the FCC phase about the $z$ axis (parallel to the $[1\bar{1} 0]$ direction), which creates some equally spaced steps or disconnections in the FCC phase at the interface [see Fig.~\ref{fig2}(c)-(e)]. As shown in the diagram, the FCC phase is tilted with respect to the ideal NW orientation relationship by an angle of $3.11^\circ$ [Fig.~\ref{fig2}(c)], $4.04^\circ$ [Fig.~\ref{fig2}(d)], and $5.77^\circ$ [Fig.~\ref{fig2}(e)]. Evidently, number of steps at the interface increases with increasing tilt angle. Further details regarding the crystallographic directions parallel and perpendicular to the BCC-FCC interface are given in Fig.~\ref{fig2} and Table~\ref{T3}. While joining the two phases, spacing between BCC and FCC region are taken to be the average of BCC (110) and FCC (111) inter-planar distance.

A schematic diagram of the disconnections present at the FCC phase [see Fig.~\ref{fig2}(c)-(e)] is presented in Fig.~\ref{fig2a}. In all the three cases, (111) forms the terrace plane (TP), but the habit plane (HP, shown by the dotted line in the figure) changes, depending on the tilt angle from ideal NW OR. For $\theta=3.11^\circ, 4.04^\circ, 5.77^\circ$, the corresponding HPs are found to be (998), (776) and (554), respectively. As shown in Fig.~\ref{fig2a}, [hkl] and $[11\bar{2}]$ direction lies in the HP and TP, respectively and the angle between these two directions is equal to the tilt angle ($\theta$) from ideal NW OR. The tilt angle ($\theta$) is also equal to the angle between the [111] direction and a vector perpendicular to the HP. $L$ denotes the length of the TP, consisting of 17, 13 and 9 atomic rows, when the HP is (998), (776) and (554), respectively. The ledge vector is marked as $\vec{l}$ and since the BCC phase is terminated by a flat inerface, $\vec{l}$ is also equal to the Burgers vector of the disconnection (defined as the sum of individual ledge vectors of the two phases).\cite{hirth1994,song2} As illustrated in Fig.~\ref{fig2}, every step is monoatomic.  The step height ($h$) is measured from the HP and it is approximately equal to 2.4 \AA.

\begin{figure}
\includegraphics[width=\linewidth]{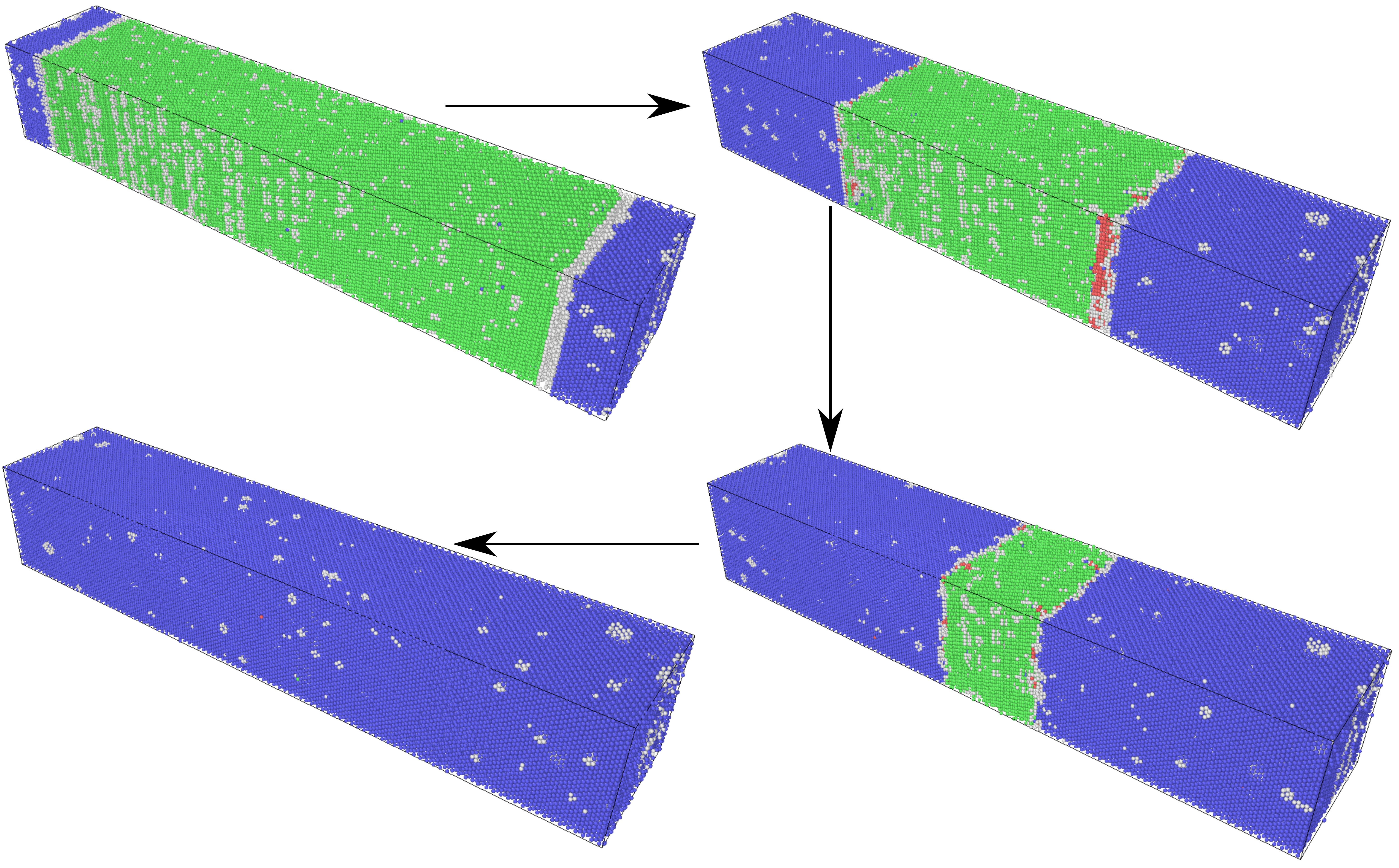}
\caption{Evolution of BCC phase (blue) at 1000 K, when the FCC phase (green) is rotated by an angel of 4.04$^\circ$ from the ideal NW OR. The snapshots are taken at 0, 5, 10 and 15 ns.} 
\label{fig3}
\end{figure}

After creating the simulation box with both the phases present in it [see Fig.~\ref{fig2}(a)], we finally run the dynamics using a NP$_x$T ensemble until the FCC phase completely transforms into the BCC phase [see Fig.~\ref{fig3}]. Depending on the temperature, it takes around 4 to 15 ns for the transformation to complete. The transformation can also be tracked by monitoring the change of potential energy, which continuously decreases as the fraction of BCC phase increases with time [see Fig.~\ref{fig4} and Fig.~\ref{fig5}].

\begin{table}
\caption{Crystallographic orientations parallel to the $x$, $y$ and $z$ direction of the simulation boxes used in this work.}
\begin{tabular}{|c|c|c|c|c|}
\hline
Phase & Direction &   Orientation   & \makecell{Size (no.of\\ atomic planes)} & Tilt angle \\
\hline
BCC   & \makecell{x \\ y \\ z}  & \makecell{ $[110]$ \\ $[\bar{1}10]$ \\ $[001]$} & \makecell{16 \\ 42 \\  54} & - \\
\hline
FCC   & \makecell{x \\ y \\ z}  & \makecell{ $[111]$ \\ $[11\bar{2}]$ \\ $[1\bar{1}0$]} & \makecell{180 \\ 37 \\ 60} & 0$^\circ$ \\
\hline
FCC &  \makecell{x \\ y \\ z}  & \makecell{ $[998]$ \\ $[44\bar{9}]$ \\ $[1\bar{1}0$]} & \makecell{180 \\ 37 \\ 60} & 3.11$^\circ$ \\
\hline
FCC &  \makecell{x \\ y \\ z}  & \makecell{ $[776]$ \\ $[33\bar{7}]$ \\ $[1\bar{1}0$]} & \makecell{180 \\ 37 \\ 60} & 4.04$^\circ$ \\
\hline
FCC &  \makecell{x \\ y \\ z}  & \makecell{ $[554]$ \\ $[22\bar{5}]$ \\ $[1\bar{1}0$]} & \makecell{180 \\ 37 \\ 60} & 5.77$^\circ$ \\
\hline
\end{tabular}
\label{T3}
\end{table}

\subsection{Interface Velocity}
\begin{figure*}
\includegraphics[width=\linewidth]{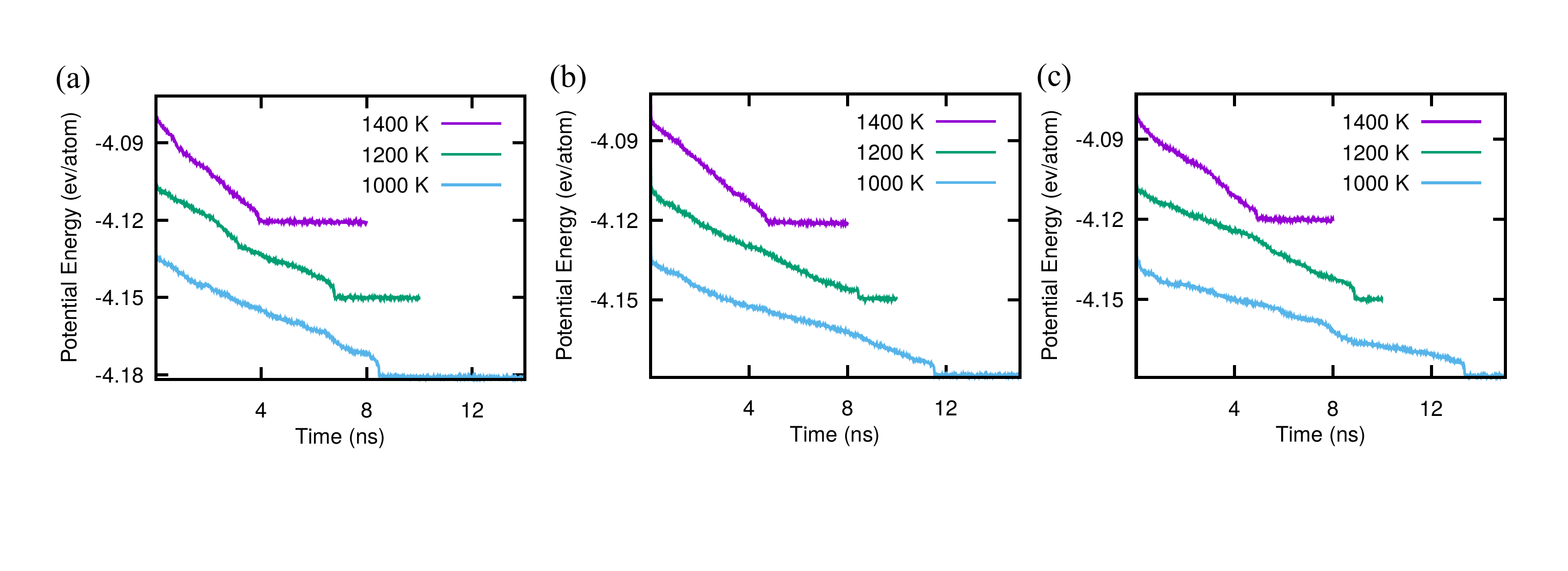}
\caption{Change of potential energy as the FCC transforms to the BCC phase at 1000, 1200 and 1400 K. The FCC phase is rotated by (a) $5.77^\circ$, (b) $4.04^\circ$ and (c) $3.11^\circ$ from the ideal NW OR. Evidently, the transformation takes lesser time at higher temperature.}
\label{fig4}
\end{figure*}

\begin{figure*}
\includegraphics[scale=1,width=\linewidth]{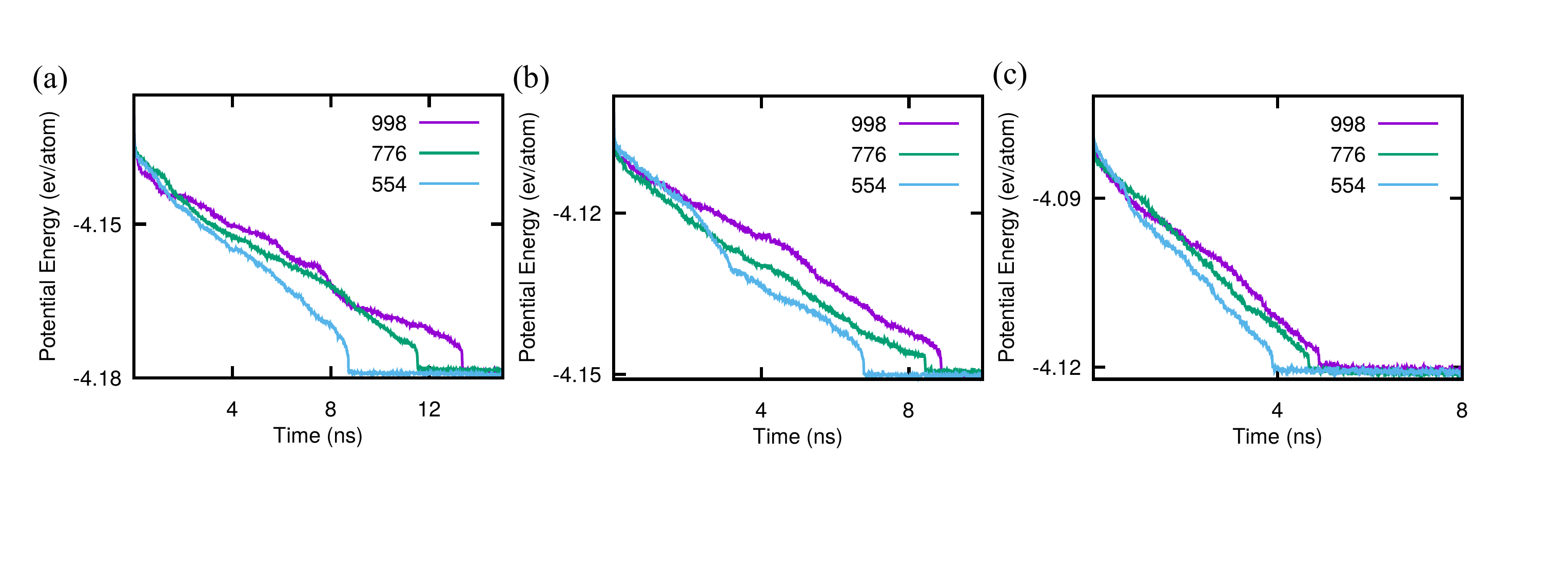}
\caption{Change of potential energy as the FCC transforms to the BCC phase at (a) 1000, (b) 1200 and (c) 1400 K. A comparison is shown among three orientations of FCC phase with respect to the ideal NW OR; 998 ($3.11^\circ$), 776 ($4.04^\circ$), 554 ($5.77^\circ$). Clearly, the transformation takes lesser time at higher angle.}
\label{fig5}
\end{figure*}

The speed ($v$) at which the austenite-ferrite interface moves can be estimated from the rate of change of potential energy $\left(\frac{dE}{dt}\right)$ during the transformation [see Fig.~\ref{fig4}] using the following equation:
\begin{equation}
v= \frac{\Omega}{2a L} \frac{dE}{dt},
\label{e2}
\end{equation}
where $a$ is the area of interface connecting the two phases, $L$ is the latent heat of solid-solid phase transformation (enthalpy difference per atom between the $\alpha$ and $\gamma$ Fe) and $\Omega$ is the volume per atom in the FCC phase. The factor 2 in the denominator takes into account the two FCC-BCC interfaces present in the simulation box. The velocity ($\vec{v}$) at which the austenite-ferrite interface moves is proportional to the driving force for the phase transition ($\Delta G_{\gamma-\alpha}$),
\begin{equation}
\vec{v} =\vec{M} \Delta G_{\gamma-\alpha},
\label{e3}
\end{equation}
where $\vec{M}$ is the interface mobility. Using the calculated values of $v$ and $\Delta G_{m}$, we further estimate the numerical value of interface mobility, which is related to the activation energy required for one atom present in the FCC phase to cross the inter-phase boundary due to thermal fluctuations and get attached to the BCC phase. The activation energy ($Q$) is calculated from the following equation
\begin{equation}
\vec{M} =\vec{M_0}\exp\left(-\frac{Q}{RT}\right),
\label{e4}
\end{equation}
where $R$ is the universal gas constant.
     
\section{Results and discussions}
\label{rd}
Among the four different BCC-FCC interfaces [see Fig.~\ref{fig2} and Table~\ref{T3}], no phase transformation is observed in case of ideal NW OR. However, when the FCC region is tilted with respect to the ideal NW OR, FCC to BCC phase transformation is indeed observed. As shown in Fig.~\ref{fig3}, growth of the BCC phase (blue) starts from both end of the simulation box and the FCC phase (green) is transformed in due course of time, ultimately converting the entire box to a BCC phase. Lack of phase transformation in case of ideal NW type interface is probably due to the absence of any defect sites, which can assist the growth of the BCC phase. On the other hand, in case of other orientations (tilted with respect to the ideal NW OR), steps or disconnections are present at the interface [see Fig.~\ref{fig1}], which is found to facilitate the growth of the BCC phase. This is going to be discussed in detail later in this section.  

Tracking the $\gamma-\alpha$ phase transformation can simply be done by monitoring the potential energy of the system as a function of time. Since $\gamma$ has higher free energy than that of $\alpha$ [see Fig.~\ref{fig1}], the potential energy of the system is going to decrease as the former is transformed into the latter phase. This is shown in Fig.~\ref{fig4} (a)-(c) for three differently oriented FCC phases at three different temperatures. For a given orientation, the phase transformation is faster at higher temperature [see Fig.~\ref{fig4} (a)-(c)]. This is because activation energy required for an atom in the $\gamma$ phase to detach from its parent FCC lattice, cross the interface and attach to the BCC lattice of the $\alpha$ phase is provided by thermal fluctuations and this process is facilitated at higher temperature.

\begin{figure}
\includegraphics[width=\linewidth]{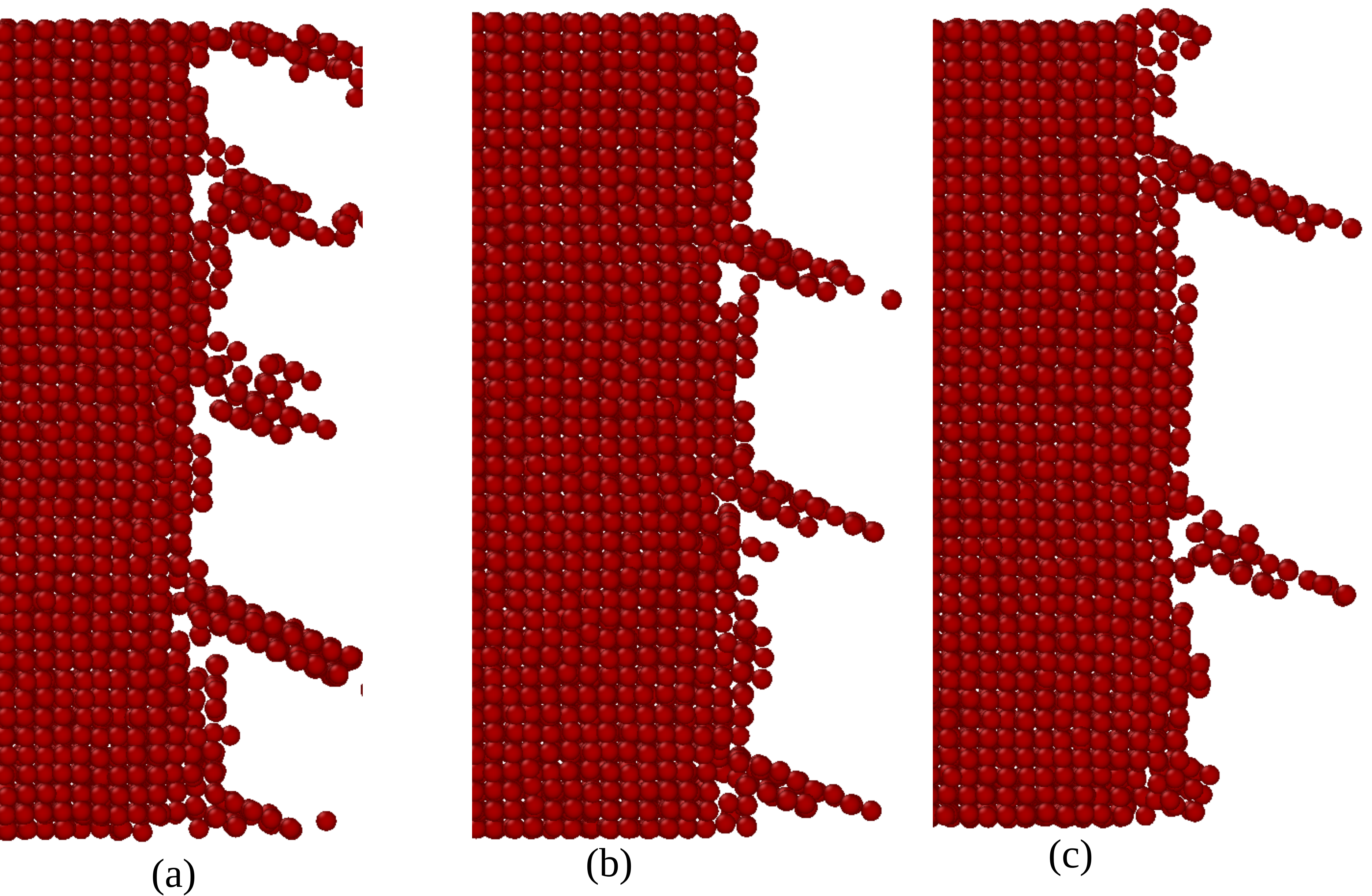}
\caption{Growth of the ferrite phase initiates at the disconnections present at the inter-phase boundary. Atoms belonging to the ferrite (BCC) phase are only shown in this figure, while the snapshot is taken at some intermediate time step during the transformation. With respect to the ideal NW OR, the austenite phase in this particular case is tilted by (a) $5.77^\circ$, (b) $4.04^\circ$ and (c) $3.11^\circ$. Depending on the number of steps or disconnections present at the interface [see Fig.~\ref{fig2}(c)-(e)], there are as many sites from which the growth of the ferrite phase starts.}
\label{fig6}
\end{figure}

It would also be interesting to compare the rate of transformation among three different orientations of the FCC phase at a given temperature. Change of potential energy as a function of time during the transformation is plotted in Fig.~\ref{fig5} (a), (b) and (c) for 1000, 1200 and 1400 K, respectively. Clearly, higher the tilt of the FCC phase with respect to the ideal NW OR, faster is the rate of transformation to the BCC phase. As already shown in Fig~\ref{fig2} (c)-(e), higher tilt angle with respect to the ideal NW OR results more steps or disconnections in the FCC side of the interface. Thus, these steps must be playing some important roles during the phase transformation process. Considering the fact that no transformation is observed in case of atomically flat ideal NW OR, as well as phase transition rate being enhanced with increasing number of steps at the interface, it appears that the steps or disconnections assist the growth of the BCC phase. This hypothesis is further confirmed by taking snapshots of the simulation box at various time steps during the transformation. Three such configurations, one each for every orientation considered in this paper, are shown in Fig.~\ref{fig6} (a)-(c), where  only the BCC phase is illustrated for the sake of visual clarity. Comparing with Fig.~\ref{fig2}(c)-(e), we conclude that depending on the number of disconnections present initially, there are as many locations from which new layers of ferrite phase starts to grow during the transformation. Note that, the disconnections present at the inter-phase boundary remain untill the whole simulation box is converted to the ferrite phase. Moreover, it is also observed that a new set of disconnections develop during the process and the interface movement takes place via the lateral motion of these disconnections. This is very similar to the process described by \citeauthor{song2}.\cite{song2}

\begin{table}
\caption{Interface velocities calculated (using Eq.~\ref{e2}) at different orientations and temperatures. These numbers are calculated by taking average of the values obtained from eight independent simulations starting with different initial velocities for each of the temperature and orientation. The unit of interface velocity is m/s.}
\begin{tabular}{|p{1.6cm}|p{2cm}|p{2cm}|p{2cm}|}
\hline
 Orientation &   T=1000 K   &   T=1200 K   & T=1400 K  \\
 \hline
 5.77$^\circ$ &1.98$\pm$0.29 &3.24$\pm$ 0.28 & 4.67$\pm$ 0.21\\
 \hline
 4.04$^\circ$ & 1.46$\pm$ 0.09 & 2.59$\pm$ 0.19 & 4.25$\pm$ 0.15\\
 \hline
 3.11$^\circ$ & 1.19$\pm$ 0.15 & 2.28$\pm$ 0.21 & 4.11$\pm$ 0.23\\
 \hline
\end{tabular}
\label{T4}
\end{table}

\begin{table}
\caption{Interface mobilities calculated (using Eq.~\ref{e3}) at different orientations and temperatures. Similar to Table~\ref{T4}, data from eight independent calculations are averaged to get the values of mobility for each of the temperature and orientation. The unit of mobility is 10$^{-3}$ m-mol/(J-s).}
\begin{tabular}{|p{1.6cm}|p{2cm}|p{2cm}|p{2cm}|}
 \hline
 Orientation &   T=1000K   &   T=1200K   & T=1400K  \\
 \hline
 5.77$^\circ$ &1.0$\pm$ 0.15 &1.8$\pm$0.16 & 2.9$\pm$0.13\\
 \hline
 4.04$^\circ$ & 0.8$\pm$0.05 & 1.5$\pm$0.11 & 2.6$\pm$0.09\\
 \hline
 3.11$^\circ$ & 0.6$\pm$0.08 & 1.3$\pm$ 0.12& 2.5$\pm$0.14\\
 \hline
\end{tabular}
\label{T5}
\end{table}

After uncovering the atomistic mechanism of austenite to ferrite phase transformation, we now estimate the speed at which the austenite-ferrite interfaces move during the transition. Interface velocity is calculated using Eq.~\ref{e2}, where $\left(\frac{dE}{dt}\right)$ is taken to be the slope obtained from a linear fit of the potential energy profiles during the transformation [see Fig.~\ref{fig4} and Fig.~\ref{fig5}]. Calculated values of interface velocity (reported in Table~\ref{T4}) lie in the range of 1.19 to 4.67 m/s, depending on the temperature and orientation of the austenite phase. Note that, interface velocities reported in Table~\ref{T4} are calculated by averaging the values obtained from eight independent simulations starting with different initial velocities for each of the temperature and orientation. Comparing with the values reported in the literature,\cite{bos,song1} interface velocities obtained in the present work are significantly lesser than that of martensitic transformation, but similar to that of massive transformation. This further confirms the transformation in the present work to be massive in nature. As expected, interface velocity for any particular orientation increases with temperature because higher thermal energy helps the atoms to cross over from the austenite to the ferrite site. Interestingly, at a given temperature, interface velocity increases as the austenite phase is tilted further away from the ideal NW OR. This is possibly because, with increasing number of steps or disconnections, there are more sites from which the growth of the ferrite phase can take place; leading to faster movement of the boundary at higher tilt angles.

\begin{figure}
\includegraphics[width=\linewidth]{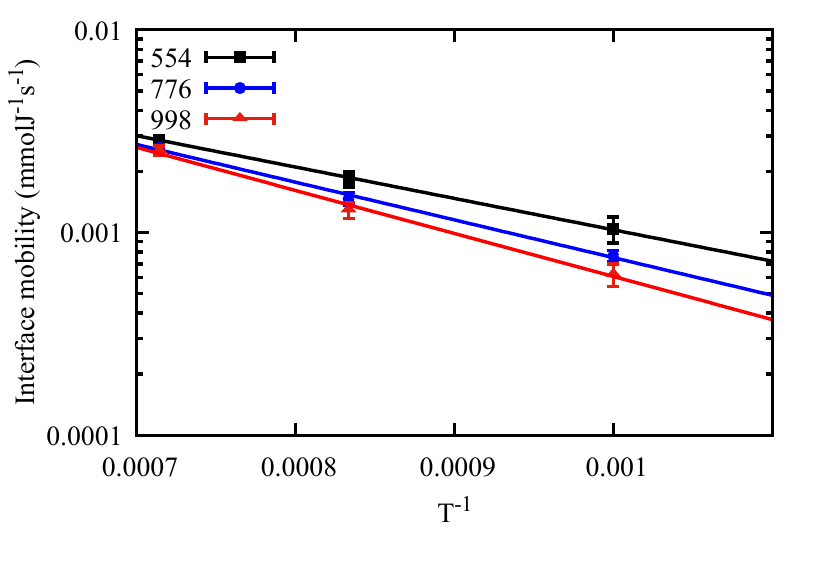}
\caption{Interface mobility plotted as a function of inverse of the absolute temperature for different orientations of the austenite phase with respect to the ideal NW OR. The slopes of the fitted lines give the value of activation energies [see Eq.~\ref{e4}].}
\label{fig7}
\end{figure}

Finally, we estimate the mobility of the interface during the austenite-ferrite transformation using Eq.~\ref{e3}. Driving force and interface velocity data are taken from Table~\ref{T2} and Table~\ref{T4}, respectively. Calculated values of mobility are reported in Table~\ref{T5}. Since interface velocity increases and $\Delta G_{\gamma-\alpha}$ decreases with temperature, mobility for a given interface orientation enhances with increasing value of $T$. Interestingly, at a particular temperature, mobility increases with the angle of tilt of the austenite phase with respect to the ideal NW OR [see Table~\ref{T5}]. This is directly related to the enhancement of the interface velocity with the tilt angle at any particular temperature, as reported in  Table~\ref{T4}. Activation energy $Q$ [see Eq.~\ref{e4}] is estimated from the slope of the $\ln M$ vs. $\frac{1}{T}$ line, as illustrated in Fig.~\ref{fig7} for all three different orientations of the austenite phase. Clearly, $Q$ decreases with increasing tilt angle from the ideal NW OR. The numerical values of the activation energy are found to be 29.62, 35.60 and 40.63 kJ/mol, when the austenite phase is tilted by 5.77$^\circ$, 4.04$^\circ$ and 3.11$^\circ$, respectively.

Comparing with experimental data,\cite{hillert,gamsjager,krielaart} calculated values of activation energies are found to be 3 to 4 times lower in our simulations, which means faster transition from the austenite to the ferrite phase. There can be several reasons behind this anomaly. Firstly, we simulate pure Fe, while most of the experiments are for Fe-C-X (where X can be Mn, Ni etc.) type of alloys and alloying elements can slow down the rate of transformation. Secondly, in our study $\alpha-\alpha$ and $\gamma-\gamma$ grain boundaries are absent. In reality, there exist a network of grain boundaries, which can hinder the mobility of $\alpha-\gamma$ inter-phase boundary. Thirdly, since we are using the Ackland potential, the driving force for the phase transition is in the higher side. In reality, driving force is very small close to the $\alpha-\gamma$ transition temperature (1185 K), which can not be captured by this particular potential. However, the atomistic mechanism of growth of the ferrite at its interface with the austenite phase is unlikely to be dependent on the choice of the empirical potential. We believe that our results are correct in this regard. 

\section{Summary and Conclusions}
\label{con}
In conclusion, we present a detailed analysis of the roles played by disconnections, which appear as steps at the $\alpha-\gamma$ inter-phase boundary, during the austenite to ferrite phase transformation. Based on calculated values of interface velocities (1.19-4.67 m/s) and mobilities (30-40 kJ/mol), we identify the mechanism of $\gamma-\alpha$ transition studied in this paper as massive transformation. We clearly show that the disconnections act as centers from which the ferrite phase starts to grow. Interestingly, higher concentration of such defects at the interface enhances the rate at which austenite transforms to ferrite. Moreover, in the absence of disconnections, atomically flat interface between $\alpha$-Fe and $\gamma$-Fe (formed according to NW ORs) remains immobile and the two solid phases coexist for the entire span (up to 20 ns) of the molecular dynamics simulations. This clearly proves that defects are crucial for the transformation to start and disconnections are certainly a type of defect which can assist in this regard. 

\section{Acknowledgements}
Authors acknowledge CC IITK for providing computational facilities by giving access to the HPC 2013. SB thanks Dr. N. P. Gurao and Dr. R. Mukherjee for useful suggestions and discussions.

\bibliography{ref}

\begin{thebibliography}{46}%
\makeatletter
\providecommand \@ifxundefined [1]{%
 \@ifx{#1\undefined}
}%
\providecommand \@ifnum [1]{%
 \ifnum #1\expandafter \@firstoftwo
 \else \expandafter \@secondoftwo
 \fi
}%
\providecommand \@ifx [1]{%
 \ifx #1\expandafter \@firstoftwo
 \else \expandafter \@secondoftwo
 \fi
}%
\providecommand \natexlab [1]{#1}%
\providecommand \enquote  [1]{``#1''}%
\providecommand \bibnamefont  [1]{#1}%
\providecommand \bibfnamefont [1]{#1}%
\providecommand \citenamefont [1]{#1}%
\providecommand \href@noop [0]{\@secondoftwo}%
\providecommand \href [0]{\begingroup \@sanitize@url \@href}%
\providecommand \@href[1]{\@@startlink{#1}\@@href}%
\providecommand \@@href[1]{\endgroup#1\@@endlink}%
\providecommand \@sanitize@url [0]{\catcode `\\12\catcode `\$12\catcode
  `\&12\catcode `\#12\catcode `\^12\catcode `\_12\catcode `\%12\relax}%
\providecommand \@@startlink[1]{}%
\providecommand \@@endlink[0]{}%
\providecommand \url  [0]{\begingroup\@sanitize@url \@url }%
\providecommand \@url [1]{\endgroup\@href {#1}{\urlprefix }}%
\providecommand \urlprefix  [0]{URL }%
\providecommand \Eprint [0]{\href }%
\providecommand \doibase [0]{http://dx.doi.org/}%
\providecommand \selectlanguage [0]{\@gobble}%
\providecommand \bibinfo  [0]{\@secondoftwo}%
\providecommand \bibfield  [0]{\@secondoftwo}%
\providecommand \translation [1]{[#1]}%
\providecommand \BibitemOpen [0]{}%
\providecommand \bibitemStop [0]{}%
\providecommand \bibitemNoStop [0]{.\EOS\space}%
\providecommand \EOS [0]{\spacefactor3000\relax}%
\providecommand \BibitemShut  [1]{\csname bibitem#1\endcsname}%
\let\auto@bib@innerbib\@empty
\bibitem [{\citenamefont {Roitburd}\ and\ \citenamefont
  {Kurdjumov}(1979)}]{roitburd}%
  \BibitemOpen
  \bibfield  {author} {\bibinfo {author} {\bibfnamefont {A.L.}\ \bibnamefont
  {Roitburd}}\ and\ \bibinfo {author} {\bibfnamefont {G.V.}\ \bibnamefont
  {Kurdjumov}},\ }\bibfield  {title} {\enquote {\bibinfo {title} {The nature of
  martensitic transformations},}\ }\href {\doibase
  https://doi.org/10.1016/0025-5416(79)90055-7} {\bibfield  {journal} {\bibinfo
   {journal} {Materials Science and Engineering}\ }\textbf {\bibinfo {volume}
  {39}},\ \bibinfo {pages} {141 -- 167} (\bibinfo {year} {1979})}\BibitemShut
  {NoStop}%
\bibitem [{\citenamefont {Hillert}\ and\ \citenamefont
  {Höglund}(2006)}]{hillert}%
  \BibitemOpen
  \bibfield  {author} {\bibinfo {author} {\bibfnamefont {M.}~\bibnamefont
  {Hillert}}\ and\ \bibinfo {author} {\bibfnamefont {L.}~\bibnamefont
  {Höglund}},\ }\bibfield  {title} {\enquote {\bibinfo {title} {Mobility of
  $\alpha/\gamma$ phase interfaces in fe alloys},}\ }\href {\doibase
  https://doi.org/10.1016/j.scriptamat.2005.12.023} {\bibfield  {journal}
  {\bibinfo  {journal} {Scripta Materialia}\ }\textbf {\bibinfo {volume}
  {54}},\ \bibinfo {pages} {1259 -- 1263} (\bibinfo {year} {2006})}\BibitemShut
  {NoStop}%
\bibitem [{\citenamefont {Zener}(1949)}]{zener}%
  \BibitemOpen
  \bibfield  {author} {\bibinfo {author} {\bibfnamefont {C.}~\bibnamefont
  {Zener}},\ }\bibfield  {title} {\enquote {\bibinfo {title} {Theory of growth
  of spherical precipitates from solid solution},}\ }\href {\doibase
  10.1063/1.1698258} {\bibfield  {journal} {\bibinfo  {journal} {Journal of
  Applied Physics}\ }\textbf {\bibinfo {volume} {20}},\ \bibinfo {pages}
  {950--953} (\bibinfo {year} {1949})}\BibitemShut {NoStop}%
\bibitem [{\citenamefont {CHRISTIAN}(2002)}]{christian}%
  \BibitemOpen
  \bibfield  {author} {\bibinfo {author} {\bibfnamefont {J.W.}\ \bibnamefont
  {CHRISTIAN}},\ }\bibfield  {title} {\enquote {\bibinfo {title} {The theory of
  transformations in metals and alloys},}\ \ }(\bibinfo  {publisher}
  {Pergamon},\ \bibinfo {address} {Oxford},\ \bibinfo {year} {2002})\ pp.\
  \bibinfo {pages} {797 -- 817}\BibitemShut {NoStop}%
\bibitem [{\citenamefont {Sietsma}\ and\ \citenamefont {van~der
  Zwaag}(2004)}]{sietsma_z}%
  \BibitemOpen
  \bibfield  {author} {\bibinfo {author} {\bibfnamefont {Jilt}\ \bibnamefont
  {Sietsma}}\ and\ \bibinfo {author} {\bibfnamefont {Sybrand}\ \bibnamefont
  {van~der Zwaag}},\ }\bibfield  {title} {\enquote {\bibinfo {title} {A concise
  model for mixed-mode phase transformations in the solid state},}\ }\href
  {\doibase https://doi.org/10.1016/j.actamat.2004.05.027} {\bibfield
  {journal} {\bibinfo  {journal} {Acta Materialia}\ }\textbf {\bibinfo {volume}
  {52}},\ \bibinfo {pages} {4143 -- 4152} (\bibinfo {year} {2004})}\BibitemShut
  {NoStop}%
\bibitem [{\citenamefont {Krielaart}\ \emph {et~al.}(1997)\citenamefont
  {Krielaart}, \citenamefont {Sietsma},\ and\ \citenamefont {Van
  Der~Zwaag}}]{krielaart}%
  \BibitemOpen
  \bibfield  {author} {\bibinfo {author} {\bibfnamefont {G.P.}\ \bibnamefont
  {Krielaart}}, \bibinfo {author} {\bibfnamefont {J.}~\bibnamefont {Sietsma}},
  \ and\ \bibinfo {author} {\bibfnamefont {S.}~\bibnamefont {Van Der~Zwaag}},\
  }\bibfield  {title} {\enquote {\bibinfo {title} {Ferrite formation in fe-c
  alloys during austenite decomposition under non-equilibrium interface
  conditions},}\ }\href {\doibase 10.1016/S0921-5093(97)00365-1} {\bibfield
  {journal} {\bibinfo  {journal} {Materials Science and Engineering A}\
  }\textbf {\bibinfo {volume} {237}},\ \bibinfo {pages} {216--223} (\bibinfo
  {year} {1997})}\BibitemShut {NoStop}%
\bibitem [{\citenamefont {Gamsjäger}\ \emph {et~al.}(2014)\citenamefont
  {Gamsjäger}, \citenamefont {Chen},\ and\ \citenamefont {Van
  Der~Zwaag}}]{gamsjager}%
  \BibitemOpen
  \bibfield  {author} {\bibinfo {author} {\bibfnamefont {E.}~\bibnamefont
  {Gamsjäger}}, \bibinfo {author} {\bibfnamefont {H.}~\bibnamefont {Chen}}, \
  and\ \bibinfo {author} {\bibfnamefont {S.}~\bibnamefont {Van Der~Zwaag}},\
  }\bibfield  {title} {\enquote {\bibinfo {title} {Application of the cyclic
  phase transformation concept for determining the effective austenite/ferrite
  interface mobility},}\ }\href {\doibase 10.1016/j.commatsci.2013.10.036}
  {\bibfield  {journal} {\bibinfo  {journal} {Computational Materials Science}\
  }\textbf {\bibinfo {volume} {83}},\ \bibinfo {pages} {92--100} (\bibinfo
  {year} {2014})}\BibitemShut {NoStop}%
\bibitem [{\citenamefont {Bain}\ and\ \citenamefont {Dunkirk}(1924)}]{bain}%
  \BibitemOpen
  \bibfield  {author} {\bibinfo {author} {\bibfnamefont {Edgar~C}\ \bibnamefont
  {Bain}}\ and\ \bibinfo {author} {\bibfnamefont {NY}~\bibnamefont {Dunkirk}},\
  }\bibfield  {title} {\enquote {\bibinfo {title} {The nature of martensite},}\
  }\href@noop {} {\bibfield  {journal} {\bibinfo  {journal} {trans. AIME}\
  }\textbf {\bibinfo {volume} {70}},\ \bibinfo {pages} {25--47} (\bibinfo
  {year} {1924})}\BibitemShut {NoStop}%
\bibitem [{\citenamefont {Nishiyama}(1934)}]{nw1934}%
  \BibitemOpen
  \bibfield  {author} {\bibinfo {author} {\bibfnamefont {Z.}~\bibnamefont
  {Nishiyama}},\ }\bibfield  {title} {\enquote {\bibinfo {title} {Mechanism of
  transformation from face-centred to body-centred cubic lattice},}\
  }\href@noop {} {\bibfield  {journal} {\bibinfo  {journal} {Sci Rep Tohoku Imp
  Univ.}\ }\textbf {\bibinfo {volume} {23}},\ \bibinfo {pages} {637–664}
  (\bibinfo {year} {1934})}\BibitemShut {NoStop}%
\bibitem [{\citenamefont {Kurdjumow}\ and\ \citenamefont
  {Sachs}(1930)}]{Kurdjumow1930}%
  \BibitemOpen
  \bibfield  {author} {\bibinfo {author} {\bibfnamefont {G.}~\bibnamefont
  {Kurdjumow}}\ and\ \bibinfo {author} {\bibfnamefont {G.}~\bibnamefont
  {Sachs}},\ }\bibfield  {title} {\enquote {\bibinfo {title} {{\"U}ber den
  mechanismus der stahlh{\"a}rtung},}\ }\href {\doibase 10.1007/BF01397346}
  {\bibfield  {journal} {\bibinfo  {journal} {Zeitschrift f{\"u}r Physik}\
  }\textbf {\bibinfo {volume} {64}},\ \bibinfo {pages} {325--343} (\bibinfo
  {year} {1930})}\BibitemShut {NoStop}%
\bibitem [{\citenamefont {Greninger}\ and\ \citenamefont {Troiano}(1949)}]{gt}%
  \BibitemOpen
  \bibfield  {author} {\bibinfo {author} {\bibfnamefont {Alden~B}\ \bibnamefont
  {Greninger}}\ and\ \bibinfo {author} {\bibfnamefont {Alexander~R}\
  \bibnamefont {Troiano}},\ }\bibfield  {title} {\enquote {\bibinfo {title}
  {The mechanism of martensite formation},}\ }\href@noop {} {\bibfield
  {journal} {\bibinfo  {journal} {JOM}\ }\textbf {\bibinfo {volume} {1}},\
  \bibinfo {pages} {590--598} (\bibinfo {year} {1949})}\BibitemShut {NoStop}%
\bibitem [{\citenamefont {Pitsch}(1959)}]{pitsch}%
  \BibitemOpen
  \bibfield  {author} {\bibinfo {author} {\bibfnamefont {W.}~\bibnamefont
  {Pitsch}},\ }\bibfield  {title} {\enquote {\bibinfo {title} {The martensite
  transformation in thin foils of iron-nitrogen alloys},}\ }\href {\doibase
  10.1080/14786435908238253} {\bibfield  {journal} {\bibinfo  {journal} {The
  Philosophical Magazine: A Journal of Theoretical Experimental and Applied
  Physics}\ }\textbf {\bibinfo {volume} {4}},\ \bibinfo {pages} {577--584}
  (\bibinfo {year} {1959})}\BibitemShut {NoStop}%
\bibitem [{\citenamefont {Fukino}\ and\ \citenamefont
  {Tsurekawa}(2008)}]{fukino}%
  \BibitemOpen
  \bibfield  {author} {\bibinfo {author} {\bibfnamefont {Tatsuya}\ \bibnamefont
  {Fukino}}\ and\ \bibinfo {author} {\bibfnamefont {Sadahiro}\ \bibnamefont
  {Tsurekawa}},\ }\bibfield  {title} {\enquote {\bibinfo {title} {In-situ
  sem/ebsd observation of $\alpha/\gamma$; phase transformation in fe-ni
  alloy},}\ }\href {\doibase 10.2320/matertrans.MAW200824} {\bibfield
  {journal} {\bibinfo  {journal} {MATERIALS TRANSACTIONS}\ }\textbf {\bibinfo
  {volume} {49}},\ \bibinfo {pages} {2770--2775} (\bibinfo {year}
  {2008})}\BibitemShut {NoStop}%
\bibitem [{\citenamefont {Ma}\ \emph {et~al.}(2017)\citenamefont {Ma},
  \citenamefont {Dudarev},\ and\ \citenamefont {Wr\'obel}}]{dudarev}%
  \BibitemOpen
  \bibfield  {author} {\bibinfo {author} {\bibfnamefont {Pui-Wai}\ \bibnamefont
  {Ma}}, \bibinfo {author} {\bibfnamefont {S.~L.}\ \bibnamefont {Dudarev}}, \
  and\ \bibinfo {author} {\bibfnamefont {Jan~S.}\ \bibnamefont {Wr\'obel}},\
  }\bibfield  {title} {\enquote {\bibinfo {title} {Dynamic simulation of
  structural phase transitions in magnetic iron},}\ }\href {\doibase
  10.1103/PhysRevB.96.094418} {\bibfield  {journal} {\bibinfo  {journal} {Phys.
  Rev. B}\ }\textbf {\bibinfo {volume} {96}},\ \bibinfo {pages} {094418}
  (\bibinfo {year} {2017})}\BibitemShut {NoStop}%
\bibitem [{\citenamefont {Katanin}\ \emph {et~al.}(2016)\citenamefont
  {Katanin}, \citenamefont {Belozerov},\ and\ \citenamefont
  {Anisimov}}]{katanin2016}%
  \BibitemOpen
  \bibfield  {author} {\bibinfo {author} {\bibfnamefont {A.~A.}\ \bibnamefont
  {Katanin}}, \bibinfo {author} {\bibfnamefont {A.~S.}\ \bibnamefont
  {Belozerov}}, \ and\ \bibinfo {author} {\bibfnamefont {V.~I.}\ \bibnamefont
  {Anisimov}},\ }\bibfield  {title} {\enquote {\bibinfo {title} {Nonlocal
  correlations in the vicinity of the
  $\ensuremath{\alpha}\ensuremath{-}\ensuremath{\gamma}$ phase transition in
  iron within a dmft plus spin-fermion model approach},}\ }\href {\doibase
  10.1103/PhysRevB.94.161117} {\bibfield  {journal} {\bibinfo  {journal} {Phys.
  Rev. B}\ }\textbf {\bibinfo {volume} {94}},\ \bibinfo {pages} {161117}
  (\bibinfo {year} {2016})}\BibitemShut {NoStop}%
\bibitem [{\citenamefont {Leonov}\ \emph {et~al.}(2011)\citenamefont {Leonov},
  \citenamefont {Poteryaev}, \citenamefont {Anisimov},\ and\ \citenamefont
  {Vollhardt}}]{leonov2011}%
  \BibitemOpen
  \bibfield  {author} {\bibinfo {author} {\bibfnamefont {I.}~\bibnamefont
  {Leonov}}, \bibinfo {author} {\bibfnamefont {A.~I.}\ \bibnamefont
  {Poteryaev}}, \bibinfo {author} {\bibfnamefont {V.~I.}\ \bibnamefont
  {Anisimov}}, \ and\ \bibinfo {author} {\bibfnamefont {D.}~\bibnamefont
  {Vollhardt}},\ }\bibfield  {title} {\enquote {\bibinfo {title} {Electronic
  correlations at the
  $\ensuremath{\alpha}\mathrm{\text{\ensuremath{-}}}\ensuremath{\gamma}$
  structural phase transition in paramagnetic iron},}\ }\href {\doibase
  10.1103/PhysRevLett.106.106405} {\bibfield  {journal} {\bibinfo  {journal}
  {Phys. Rev. Lett.}\ }\textbf {\bibinfo {volume} {106}},\ \bibinfo {pages}
  {106405} (\bibinfo {year} {2011})}\BibitemShut {NoStop}%
\bibitem [{\citenamefont {Leonov}\ \emph {et~al.}(2012)\citenamefont {Leonov},
  \citenamefont {Poteryaev}, \citenamefont {Anisimov},\ and\ \citenamefont
  {Vollhardt}}]{leonov2012}%
  \BibitemOpen
  \bibfield  {author} {\bibinfo {author} {\bibfnamefont {I.}~\bibnamefont
  {Leonov}}, \bibinfo {author} {\bibfnamefont {A.~I.}\ \bibnamefont
  {Poteryaev}}, \bibinfo {author} {\bibfnamefont {V.~I.}\ \bibnamefont
  {Anisimov}}, \ and\ \bibinfo {author} {\bibfnamefont {D.}~\bibnamefont
  {Vollhardt}},\ }\bibfield  {title} {\enquote {\bibinfo {title} {Calculated
  phonon spectra of paramagnetic iron at the
  $\ensuremath{\alpha}$-$\ensuremath{\gamma}$ phase transition},}\ }\href
  {\doibase 10.1103/PhysRevB.85.020401} {\bibfield  {journal} {\bibinfo
  {journal} {Phys. Rev. B}\ }\textbf {\bibinfo {volume} {85}},\ \bibinfo
  {pages} {020401} (\bibinfo {year} {2012})}\BibitemShut {NoStop}%
\bibitem [{\citenamefont {Ou}(2017)}]{ou}%
  \BibitemOpen
  \bibfield  {author} {\bibinfo {author} {\bibfnamefont {X.}~\bibnamefont
  {Ou}},\ }\bibfield  {title} {\enquote {\bibinfo {title} {Molecular dynamics
  simulations of fcc-to-bcc transformation in pure iron: a review},}\ }\href
  {\doibase 10.1080/02670836.2016.1204064} {\bibfield  {journal} {\bibinfo
  {journal} {Materials Science and Technology}\ }\textbf {\bibinfo {volume}
  {33}},\ \bibinfo {pages} {822--835} (\bibinfo {year} {2017})}\BibitemShut
  {NoStop}%
\bibitem [{\citenamefont {Ou}\ \emph {et~al.}(2016)\citenamefont {Ou},
  \citenamefont {Sietsma},\ and\ \citenamefont {Santofimia}}]{ou_w}%
  \BibitemOpen
  \bibfield  {author} {\bibinfo {author} {\bibfnamefont {X}~\bibnamefont {Ou}},
  \bibinfo {author} {\bibfnamefont {J}~\bibnamefont {Sietsma}}, \ and\ \bibinfo
  {author} {\bibfnamefont {M~J}\ \bibnamefont {Santofimia}},\ }\bibfield
  {title} {\enquote {\bibinfo {title} {Molecular dynamics simulations of the
  mechanisms controlling the propagation of bcc/fcc semi-coherent interfaces in
  iron},}\ }\href {http://stacks.iop.org/0965-0393/24/i=5/a=055019} {\bibfield
  {journal} {\bibinfo  {journal} {Modelling and Simulation in Materials Science
  and Engineering}\ }\textbf {\bibinfo {volume} {24}},\ \bibinfo {pages}
  {055019} (\bibinfo {year} {2016})}\BibitemShut {NoStop}%
\bibitem [{\citenamefont {Bos}\ \emph {et~al.}(2006)\citenamefont {Bos},
  \citenamefont {Sietsma},\ and\ \citenamefont {Thijsse}}]{bos}%
  \BibitemOpen
  \bibfield  {author} {\bibinfo {author} {\bibfnamefont {C.}~\bibnamefont
  {Bos}}, \bibinfo {author} {\bibfnamefont {J.}~\bibnamefont {Sietsma}}, \ and\
  \bibinfo {author} {\bibfnamefont {B.~J.}\ \bibnamefont {Thijsse}},\
  }\bibfield  {title} {\enquote {\bibinfo {title} {Molecular dynamics
  simulation of interface dynamics during the fcc-bcc transformation of a
  martensitic nature},}\ }\href {\doibase 10.1103/PhysRevB.73.104117}
  {\bibfield  {journal} {\bibinfo  {journal} {Phys. Rev. B}\ }\textbf {\bibinfo
  {volume} {73}},\ \bibinfo {pages} {104117} (\bibinfo {year}
  {2006})}\BibitemShut {NoStop}%
\bibitem [{\citenamefont {Song}\ and\ \citenamefont {Hoyt}(2012)}]{song1}%
  \BibitemOpen
  \bibfield  {author} {\bibinfo {author} {\bibfnamefont {H.}~\bibnamefont
  {Song}}\ and\ \bibinfo {author} {\bibfnamefont {J.J.}\ \bibnamefont {Hoyt}},\
  }\bibfield  {title} {\enquote {\bibinfo {title} {A molecular dynamics
  simulation study of the velocities, mobility and activation energy of an
  austenite–ferrite interface in pure fe},}\ }\href {\doibase
  https://doi.org/10.1016/j.actamat.2012.04.023} {\bibfield  {journal}
  {\bibinfo  {journal} {Acta Materialia}\ }\textbf {\bibinfo {volume} {60}},\
  \bibinfo {pages} {4328 -- 4335} (\bibinfo {year} {2012})}\BibitemShut
  {NoStop}%
\bibitem [{\citenamefont {Song}\ and\ \citenamefont {Hoyt}(2013)}]{song2}%
  \BibitemOpen
  \bibfield  {author} {\bibinfo {author} {\bibfnamefont {H.}~\bibnamefont
  {Song}}\ and\ \bibinfo {author} {\bibfnamefont {J.J.}\ \bibnamefont {Hoyt}},\
  }\bibfield  {title} {\enquote {\bibinfo {title} {An atomistic simulation
  study of the migration of an austenite–ferrite interface in pure fe},}\
  }\href {\doibase https://doi.org/10.1016/j.actamat.2012.10.028} {\bibfield
  {journal} {\bibinfo  {journal} {Acta Materialia}\ }\textbf {\bibinfo {volume}
  {61}},\ \bibinfo {pages} {1189 -- 1196} (\bibinfo {year} {2013})}\BibitemShut
  {NoStop}%
\bibitem [{\citenamefont {Tateyama}\ \emph {et~al.}(2008)\citenamefont
  {Tateyama}, \citenamefont {Shibuta},\ and\ \citenamefont
  {Suzuki}}]{tateyama}%
  \BibitemOpen
  \bibfield  {author} {\bibinfo {author} {\bibfnamefont {Shinji}\ \bibnamefont
  {Tateyama}}, \bibinfo {author} {\bibfnamefont {Yasushi}\ \bibnamefont
  {Shibuta}}, \ and\ \bibinfo {author} {\bibfnamefont {Toshio}\ \bibnamefont
  {Suzuki}},\ }\bibfield  {title} {\enquote {\bibinfo {title} {A molecular
  dynamics study of the fcc–bcc phase transformation kinetics of iron},}\
  }\href {\doibase https://doi.org/10.1016/j.scriptamat.2008.06.054} {\bibfield
   {journal} {\bibinfo  {journal} {Scripta Materialia}\ }\textbf {\bibinfo
  {volume} {59}},\ \bibinfo {pages} {971 -- 974} (\bibinfo {year}
  {2008})}\BibitemShut {NoStop}%
\bibitem [{\citenamefont {Wang}\ and\ \citenamefont
  {Urbassek}(2013)}]{urbassek}%
  \BibitemOpen
  \bibfield  {author} {\bibinfo {author} {\bibfnamefont {Binjun}\ \bibnamefont
  {Wang}}\ and\ \bibinfo {author} {\bibfnamefont {Herbert~M.}\ \bibnamefont
  {Urbassek}},\ }\bibfield  {title} {\enquote {\bibinfo {title} {Phase
  transitions in an fe system containing a bcc/fcc phase boundary: An atomistic
  study},}\ }\href {\doibase 10.1103/PhysRevB.87.104108} {\bibfield  {journal}
  {\bibinfo  {journal} {Phys. Rev. B}\ }\textbf {\bibinfo {volume} {87}},\
  \bibinfo {pages} {104108} (\bibinfo {year} {2013})}\BibitemShut {NoStop}%
\bibitem [{\citenamefont {Karewar}\ \emph {et~al.}(2018)\citenamefont
  {Karewar}, \citenamefont {Sietsma},\ and\ \citenamefont
  {Santofimia}}]{sietsma_acta18}%
  \BibitemOpen
  \bibfield  {author} {\bibinfo {author} {\bibfnamefont {S.}~\bibnamefont
  {Karewar}}, \bibinfo {author} {\bibfnamefont {J.}~\bibnamefont {Sietsma}}, \
  and\ \bibinfo {author} {\bibfnamefont {M.J.}\ \bibnamefont {Santofimia}},\
  }\bibfield  {title} {\enquote {\bibinfo {title} {Effect of pre-existing
  defects in the parent fcc phase on atomistic mechanisms during the
  martensitic transformation in pure fe: A molecular dynamics study},}\ }\href
  {\doibase https://doi.org/10.1016/j.actamat.2017.09.049} {\bibfield
  {journal} {\bibinfo  {journal} {Acta Materialia}\ }\textbf {\bibinfo {volume}
  {142}},\ \bibinfo {pages} {71 -- 81} (\bibinfo {year} {2018})}\BibitemShut
  {NoStop}%
\bibitem [{\citenamefont {Hirth}(1994)}]{hirth1994}%
  \BibitemOpen
  \bibfield  {author} {\bibinfo {author} {\bibfnamefont {J.P.}\ \bibnamefont
  {Hirth}},\ }\bibfield  {title} {\enquote {\bibinfo {title} {Dislocations,
  steps and disconnections at interfaces},}\ }\href {\doibase
  https://doi.org/10.1016/0022-3697(94)90118-X} {\bibfield  {journal} {\bibinfo
   {journal} {Journal of Physics and Chemistry of Solids}\ }\textbf {\bibinfo
  {volume} {55}},\ \bibinfo {pages} {985 -- 989} (\bibinfo {year}
  {1994})}\BibitemShut {NoStop}%
\bibitem [{\citenamefont {Hirth}\ and\ \citenamefont
  {Pond}(1996)}]{hirth1996acta}%
  \BibitemOpen
  \bibfield  {author} {\bibinfo {author} {\bibfnamefont {J.P.}\ \bibnamefont
  {Hirth}}\ and\ \bibinfo {author} {\bibfnamefont {R.C.}\ \bibnamefont
  {Pond}},\ }\bibfield  {title} {\enquote {\bibinfo {title} {Steps,
  dislocations and disconnections as interface defects relating to structure
  and phase transformations},}\ }\href {\doibase
  https://doi.org/10.1016/S1359-6454(96)00132-2} {\bibfield  {journal}
  {\bibinfo  {journal} {Acta Materialia}\ }\textbf {\bibinfo {volume} {44}},\
  \bibinfo {pages} {4749 -- 4763} (\bibinfo {year} {1996})}\BibitemShut
  {NoStop}%
\bibitem [{\citenamefont {Howe}\ \emph {et~al.}(2009)\citenamefont {Howe},
  \citenamefont {Pond},\ and\ \citenamefont {Hirth}}]{hirth_review}%
  \BibitemOpen
  \bibfield  {author} {\bibinfo {author} {\bibfnamefont {J.M.}\ \bibnamefont
  {Howe}}, \bibinfo {author} {\bibfnamefont {R.C.}\ \bibnamefont {Pond}}, \
  and\ \bibinfo {author} {\bibfnamefont {J.P.}\ \bibnamefont {Hirth}},\
  }\bibfield  {title} {\enquote {\bibinfo {title} {The role of disconnections
  in phase transformations},}\ }\href {\doibase
  https://doi.org/10.1016/j.pmatsci.2009.04.001} {\bibfield  {journal}
  {\bibinfo  {journal} {Progress in Materials Science}\ }\textbf {\bibinfo
  {volume} {54}},\ \bibinfo {pages} {792 -- 838} (\bibinfo {year}
  {2009})}\BibitemShut {NoStop}%
\bibitem [{\citenamefont {Zhang}\ \emph {et~al.}(2016)\citenamefont {Zhang},
  \citenamefont {Gu},\ and\ \citenamefont {Dai}}]{zhang}%
  \BibitemOpen
  \bibfield  {author} {\bibinfo {author} {\bibfnamefont {W.-Z.}\ \bibnamefont
  {Zhang}}, \bibinfo {author} {\bibfnamefont {X.-F.}\ \bibnamefont {Gu}}, \
  and\ \bibinfo {author} {\bibfnamefont {F.-Z.}\ \bibnamefont {Dai}},\
  }\bibfield  {title} {\enquote {\bibinfo {title} {Faceted interfaces: A key
  feature to quantitative understanding of transformation morphology},}\ }\href
  {\doibase 10.1038/npjcompumats.2016.21} {\bibfield  {journal} {\bibinfo
  {journal} {npj Computational Materials}\ }\textbf {\bibinfo {volume} {2}},\
  \bibinfo {pages} {16021} (\bibinfo {year} {2016})}\BibitemShut {NoStop}%
\bibitem [{\citenamefont {Maresca}\ and\ \citenamefont
  {Curtin}(2017)}]{maresca_acta18}%
  \BibitemOpen
  \bibfield  {author} {\bibinfo {author} {\bibfnamefont {F.}~\bibnamefont
  {Maresca}}\ and\ \bibinfo {author} {\bibfnamefont {W.A.}\ \bibnamefont
  {Curtin}},\ }\bibfield  {title} {\enquote {\bibinfo {title} {The
  austenite/lath martensite interface in steels: Structure, athermal motion,
  and in-situ transformation strain revealed by simulation and theory},}\
  }\href {\doibase https://doi.org/10.1016/j.actamat.2017.05.044} {\bibfield
  {journal} {\bibinfo  {journal} {Acta Materialia}\ }\textbf {\bibinfo {volume}
  {134}},\ \bibinfo {pages} {302 -- 323} (\bibinfo {year} {2017})}\BibitemShut
  {NoStop}%
\bibitem [{\citenamefont {Zheng}\ \emph {et~al.}(2018)\citenamefont {Zheng},
  \citenamefont {Williams}, \citenamefont {Viswanathan}, \citenamefont
  {Clark},\ and\ \citenamefont {Fraser}}]{Ti_ref2}%
  \BibitemOpen
  \bibfield  {author} {\bibinfo {author} {\bibfnamefont {Yufeng}\ \bibnamefont
  {Zheng}}, \bibinfo {author} {\bibfnamefont {Robert~E.A.}\ \bibnamefont
  {Williams}}, \bibinfo {author} {\bibfnamefont {Gopal~B.}\ \bibnamefont
  {Viswanathan}}, \bibinfo {author} {\bibfnamefont {William~A.T.}\ \bibnamefont
  {Clark}}, \ and\ \bibinfo {author} {\bibfnamefont {Hamish~L.}\ \bibnamefont
  {Fraser}},\ }\bibfield  {title} {\enquote {\bibinfo {title} {Determination of
  the structure of $\alpha-\beta$ interfaces in metastable $\beta$-ti
  alloys},}\ }\href {\doibase https://doi.org/10.1016/j.actamat.2018.03.003}
  {\bibfield  {journal} {\bibinfo  {journal} {Acta Materialia}\ }\textbf
  {\bibinfo {volume} {150}},\ \bibinfo {pages} {25 -- 39} (\bibinfo {year}
  {2018})}\BibitemShut {NoStop}%
\bibitem [{\citenamefont {Howe}\ \emph {et~al.}(2002)\citenamefont {Howe},
  \citenamefont {Reynolds},\ and\ \citenamefont {Vasudevan}}]{howe}%
  \BibitemOpen
  \bibfield  {author} {\bibinfo {author} {\bibfnamefont {James~M.}\
  \bibnamefont {Howe}}, \bibinfo {author} {\bibfnamefont {William~T.}\
  \bibnamefont {Reynolds}}, \ and\ \bibinfo {author} {\bibfnamefont {Vijay~K.}\
  \bibnamefont {Vasudevan}},\ }\bibfield  {title} {\enquote {\bibinfo {title}
  {Static and in-situ high-resolution transmission electron microscopy
  investigations of the atomic structure and dynamics of massive transformation
  interfaces in a ti-al alloy},}\ }\href {\doibase 10.1007/s11661-002-0362-4}
  {\bibfield  {journal} {\bibinfo  {journal} {Metallurgical and Materials
  Transactions A}\ }\textbf {\bibinfo {volume} {33}},\ \bibinfo {pages}
  {2391--2411} (\bibinfo {year} {2002})}\BibitemShut {NoStop}%
\bibitem [{\citenamefont {Hall}\ \emph {et~al.}(1972)\citenamefont {Hall},
  \citenamefont {Aaronson},\ and\ \citenamefont {Kinsma}}]{hall}%
  \BibitemOpen
  \bibfield  {author} {\bibinfo {author} {\bibfnamefont {M.G.}\ \bibnamefont
  {Hall}}, \bibinfo {author} {\bibfnamefont {H.I.}\ \bibnamefont {Aaronson}}, \
  and\ \bibinfo {author} {\bibfnamefont {K.R.}\ \bibnamefont {Kinsma}},\
  }\bibfield  {title} {\enquote {\bibinfo {title} {The structure of nearly
  coherent fcc: bcc boundaries in a cucr alloy},}\ }\href {\doibase
  https://doi.org/10.1016/0039-6028(72)90264-6} {\bibfield  {journal} {\bibinfo
   {journal} {Surface Science}\ }\textbf {\bibinfo {volume} {31}},\ \bibinfo
  {pages} {257 -- 274} (\bibinfo {year} {1972})}\BibitemShut {NoStop}%
\bibitem [{\citenamefont {Rigsbee}\ and\ \citenamefont
  {Aaronson}(1979)}]{aaronson}%
  \BibitemOpen
  \bibfield  {author} {\bibinfo {author} {\bibfnamefont {J.M}\ \bibnamefont
  {Rigsbee}}\ and\ \bibinfo {author} {\bibfnamefont {H.I}\ \bibnamefont
  {Aaronson}},\ }\bibfield  {title} {\enquote {\bibinfo {title} {A computer
  modeling study of partially coherent f.c.c.:b.c.c. boundaries},}\ }\href
  {\doibase https://doi.org/10.1016/0001-6160(79)90028-2} {\bibfield  {journal}
  {\bibinfo  {journal} {Acta Metallurgica}\ }\textbf {\bibinfo {volume} {27}},\
  \bibinfo {pages} {351 -- 363} (\bibinfo {year} {1979})}\BibitemShut {NoStop}%
\bibitem [{\citenamefont {Shiflet}\ and\ \citenamefont {van~der
  Merwe}(1994)}]{shiflet}%
  \BibitemOpen
  \bibfield  {author} {\bibinfo {author} {\bibfnamefont {G.J.}\ \bibnamefont
  {Shiflet}}\ and\ \bibinfo {author} {\bibfnamefont {J.H.}\ \bibnamefont
  {van~der Merwe}},\ }\bibfield  {title} {\enquote {\bibinfo {title} {The role
  of structural ledges at phase boundaries—ii. f.c.c.-b.c.c. interfaces in
  nishiyama-wasserman orientation},}\ }\href {\doibase
  https://doi.org/10.1016/0956-7151(94)90135-X} {\bibfield  {journal} {\bibinfo
   {journal} {Acta Metallurgica et Materialia}\ }\textbf {\bibinfo {volume}
  {42}},\ \bibinfo {pages} {1189 -- 1198} (\bibinfo {year} {1994})}\BibitemShut
  {NoStop}%
\bibitem [{\citenamefont {Moritani}\ \emph {et~al.}(2002)\citenamefont
  {Moritani}, \citenamefont {Miyajima}, \citenamefont {Furuhara},\ and\
  \citenamefont {Maki}}]{moritani}%
  \BibitemOpen
  \bibfield  {author} {\bibinfo {author} {\bibfnamefont {T.}~\bibnamefont
  {Moritani}}, \bibinfo {author} {\bibfnamefont {N.}~\bibnamefont {Miyajima}},
  \bibinfo {author} {\bibfnamefont {T.}~\bibnamefont {Furuhara}}, \ and\
  \bibinfo {author} {\bibfnamefont {T.}~\bibnamefont {Maki}},\ }\bibfield
  {title} {\enquote {\bibinfo {title} {Comparison of interphase boundary
  structure between bainite and martensite in steel},}\ }\href {\doibase
  https://doi.org/10.1016/S1359-6462(02)00128-8} {\bibfield  {journal}
  {\bibinfo  {journal} {Scripta Materialia}\ }\textbf {\bibinfo {volume}
  {47}},\ \bibinfo {pages} {193 -- 199} (\bibinfo {year} {2002})}\BibitemShut
  {NoStop}%
\bibitem [{lam(1995)}]{lammps}%
  \BibitemOpen
  \bibfield  {title} {\enquote {\bibinfo {title} {Fast parallel algorithms for
  short-range molecular dynamics},}\ }\href
  {http://dx.doi.org/10.1006/jcph.1995.1039} {\bibfield  {journal} {\bibinfo
  {journal} {J. Comput. Phys.}\ }\textbf {\bibinfo {volume} {117}},\ \bibinfo
  {pages} {1} (\bibinfo {year} {1995})},\ \bibinfo {note}
  {http://lammps.sandia.gov}\BibitemShut {NoStop}%
\bibitem [{\citenamefont {Ackland}\ \emph {et~al.}(1997)\citenamefont
  {Ackland}, \citenamefont {Bacon}, \citenamefont {Calder},\ and\ \citenamefont
  {Harry}}]{ackland}%
  \BibitemOpen
  \bibfield  {author} {\bibinfo {author} {\bibfnamefont {G.~J.}\ \bibnamefont
  {Ackland}}, \bibinfo {author} {\bibfnamefont {D.~J.}\ \bibnamefont {Bacon}},
  \bibinfo {author} {\bibfnamefont {A.~F.}\ \bibnamefont {Calder}}, \ and\
  \bibinfo {author} {\bibfnamefont {T.}~\bibnamefont {Harry}},\ }\bibfield
  {title} {\enquote {\bibinfo {title} {Computer simulation of point defect
  properties in dilute fe-cu alloy using a many-body interatomic potential},}\
  }\href {\doibase 10.1080/01418619708207198} {\bibfield  {journal} {\bibinfo
  {journal} {Philosophical Magazine A}\ }\textbf {\bibinfo {volume} {75}},\
  \bibinfo {pages} {713--732} (\bibinfo {year} {1997})}\BibitemShut {NoStop}%
\bibitem [{\citenamefont {Sun}\ \emph {et~al.}(2004)\citenamefont {Sun},
  \citenamefont {Asta}, \citenamefont {Hoyt}, \citenamefont {Mendelev},\ and\
  \citenamefont {Srolovitz}}]{sunprb2004}%
  \BibitemOpen
  \bibfield  {author} {\bibinfo {author} {\bibfnamefont {D.~Y.}\ \bibnamefont
  {Sun}}, \bibinfo {author} {\bibfnamefont {M.}~\bibnamefont {Asta}}, \bibinfo
  {author} {\bibfnamefont {J.~J.}\ \bibnamefont {Hoyt}}, \bibinfo {author}
  {\bibfnamefont {M.~I.}\ \bibnamefont {Mendelev}}, \ and\ \bibinfo {author}
  {\bibfnamefont {D.~J.}\ \bibnamefont {Srolovitz}},\ }\bibfield  {title}
  {\enquote {\bibinfo {title} {Crystal-melt interfacial free energies in
  metals: fcc versus bcc},}\ }\href {\doibase 10.1103/PhysRevB.69.020102}
  {\bibfield  {journal} {\bibinfo  {journal} {Phys. Rev. B}\ }\textbf {\bibinfo
  {volume} {69}},\ \bibinfo {pages} {020102} (\bibinfo {year}
  {2004})}\BibitemShut {NoStop}%
\bibitem [{\citenamefont {Abe}\ \emph {et~al.}(2016)\citenamefont {Abe},
  \citenamefont {Tsuru}, \citenamefont {Shi}, \citenamefont {Oono},\ and\
  \citenamefont {Ukai}}]{ack_ref}%
  \BibitemOpen
  \bibfield  {author} {\bibinfo {author} {\bibfnamefont {Yosuke}\ \bibnamefont
  {Abe}}, \bibinfo {author} {\bibfnamefont {Tomohito}\ \bibnamefont {Tsuru}},
  \bibinfo {author} {\bibfnamefont {Shi}\ \bibnamefont {Shi}}, \bibinfo
  {author} {\bibfnamefont {Naoko}\ \bibnamefont {Oono}}, \ and\ \bibinfo
  {author} {\bibfnamefont {Shigeharu}\ \bibnamefont {Ukai}},\ }\bibfield
  {title} {\enquote {\bibinfo {title} {Effect of the dilation caused by helium
  bubbles on edge dislocation motion in α-iron: molecular dynamics
  simulation},}\ }\href {\doibase 10.1080/00223131.2015.1133332} {\bibfield
  {journal} {\bibinfo  {journal} {Journal of Nuclear Science and Technology}\
  }\textbf {\bibinfo {volume} {53}},\ \bibinfo {pages} {1528--1534} (\bibinfo
  {year} {2016})}\BibitemShut {NoStop}%
\bibitem [{\citenamefont {Terentyev}\ \emph {et~al.}(2016)\citenamefont
  {Terentyev}, \citenamefont {Bakaev}, \citenamefont {Neck},\ and\
  \citenamefont {Zhurkin}}]{ack_ref2}%
  \BibitemOpen
  \bibfield  {author} {\bibinfo {author} {\bibfnamefont {D.}~\bibnamefont
  {Terentyev}}, \bibinfo {author} {\bibfnamefont {A.}~\bibnamefont {Bakaev}},
  \bibinfo {author} {\bibfnamefont {D.~Van}\ \bibnamefont {Neck}}, \ and\
  \bibinfo {author} {\bibfnamefont {E.~E.}\ \bibnamefont {Zhurkin}},\
  }\bibfield  {title} {\enquote {\bibinfo {title} {Glide of dislocations in $<1
  1 1>\{3 2 1\}$ slip system: an atomistic study},}\ }\href {\doibase
  10.1080/14786435.2015.1126369} {\bibfield  {journal} {\bibinfo  {journal}
  {Philosophical Magazine}\ }\textbf {\bibinfo {volume} {96}},\ \bibinfo
  {pages} {71--83} (\bibinfo {year} {2016})}\BibitemShut {NoStop}%
\bibitem [{\citenamefont {Mendelev}\ \emph {et~al.}(2003)\citenamefont
  {Mendelev}, \citenamefont {Han}, \citenamefont {Srolovitz}, \citenamefont
  {Ackland}, \citenamefont {Sun},\ and\ \citenamefont {Asta}}]{mendelev}%
  \BibitemOpen
  \bibfield  {author} {\bibinfo {author} {\bibfnamefont {M.~I.}\ \bibnamefont
  {Mendelev}}, \bibinfo {author} {\bibfnamefont {S.}~\bibnamefont {Han}},
  \bibinfo {author} {\bibfnamefont {D.~J.}\ \bibnamefont {Srolovitz}}, \bibinfo
  {author} {\bibfnamefont {G.~J.}\ \bibnamefont {Ackland}}, \bibinfo {author}
  {\bibfnamefont {D.~Y.}\ \bibnamefont {Sun}}, \ and\ \bibinfo {author}
  {\bibfnamefont {M.}~\bibnamefont {Asta}},\ }\bibfield  {title} {\enquote
  {\bibinfo {title} {Development of new interatomic potentials appropriate for
  crystalline and liquid iron},}\ }\href {\doibase
  10.1080/14786430310001613264} {\bibfield  {journal} {\bibinfo  {journal}
  {Philosophical Magazine}\ }\textbf {\bibinfo {volume} {83}},\ \bibinfo
  {pages} {3977--3994} (\bibinfo {year} {2003})}\BibitemShut {NoStop}%
\bibitem [{\citenamefont {Hirth}\ and\ \citenamefont {Lothe}(1982)}]{hl82}%
  \BibitemOpen
  \bibfield  {author} {\bibinfo {author} {\bibfnamefont {John~P.}\ \bibnamefont
  {Hirth}}\ and\ \bibinfo {author} {\bibfnamefont {Jens}\ \bibnamefont
  {Lothe}},\ }\href@noop {} {\emph {\bibinfo {title} {{Theory of
  Dislocations}}}},\ \bibinfo {edition} {2nd}\ ed.\ (\bibinfo  {publisher}
  {John Wiley \& Sons},\ \bibinfo {address} {New York},\ \bibinfo {year}
  {1982})\BibitemShut {NoStop}%
\bibitem [{\citenamefont {Morris}\ and\ \citenamefont {Song}(2002)}]{morris}%
  \BibitemOpen
  \bibfield  {author} {\bibinfo {author} {\bibfnamefont {James~R.}\
  \bibnamefont {Morris}}\ and\ \bibinfo {author} {\bibfnamefont {Xueyu}\
  \bibnamefont {Song}},\ }\bibfield  {title} {\enquote {\bibinfo {title} {The
  melting lines of model systems calculated from coexistence simulations},}\
  }\href {\doibase 10.1063/1.1474581} {\bibfield  {journal} {\bibinfo
  {journal} {The Journal of Chemical Physics}\ }\textbf {\bibinfo {volume}
  {116}},\ \bibinfo {pages} {9352--9358} (\bibinfo {year} {2002})}\BibitemShut
  {NoStop}%
\bibitem [{\citenamefont {Kaufman}\ \emph {et~al.}(1963)\citenamefont
  {Kaufman}, \citenamefont {Clougherty},\ and\ \citenamefont
  {Weiss}}]{KAUFMAN1963323}%
  \BibitemOpen
  \bibfield  {author} {\bibinfo {author} {\bibfnamefont {Larry}\ \bibnamefont
  {Kaufman}}, \bibinfo {author} {\bibfnamefont {E.V}\ \bibnamefont
  {Clougherty}}, \ and\ \bibinfo {author} {\bibfnamefont {R.J}\ \bibnamefont
  {Weiss}},\ }\bibfield  {title} {\enquote {\bibinfo {title} {The lattice
  stability of metals—iii. iron},}\ }\href {\doibase
  https://doi.org/10.1016/0001-6160(63)90157-3} {\bibfield  {journal} {\bibinfo
   {journal} {Acta Metallurgica}\ }\textbf {\bibinfo {volume} {11}},\ \bibinfo
  {pages} {323 -- 335} (\bibinfo {year} {1963})}\BibitemShut {NoStop}%
\bibitem [{\citenamefont {Stukowski}(2010)}]{ovito}%
  \BibitemOpen
  \bibfield  {author} {\bibinfo {author} {\bibfnamefont {Alexander}\
  \bibnamefont {Stukowski}},\ }\bibfield  {title} {\enquote {\bibinfo {title}
  {Visualization and analysis of atomistic simulation data with ovito, the open
  visualization tool},}\ }\href
  {http://stacks.iop.org/0965-0393/18/i=1/a=015012} {\bibfield  {journal}
  {\bibinfo  {journal} {Modelling and Simulation in Materials Science and
  Engineering}\ }\textbf {\bibinfo {volume} {18}},\ \bibinfo {pages} {015012}
  (\bibinfo {year} {2010})},\ \bibinfo {note} {http://ovito.org/}\BibitemShut
  {NoStop}%
\end{thebibliography}%
\end{document}